\documentclass[useAMS,usenatbib]{mn2e}
\def\revtex@jnl{AAS}
\usepackage{amssymb,amsmath}
\usepackage{longtable}
\usepackage{verbatim}
\usepackage{graphics,graphicx}
\usepackage{natbib}
\usepackage{rotating}
\usepackage{epsfig}
\usepackage[toc,page]{appendix}
\usepackage{multirow}
\usepackage{placeins}

\newcommand{\chandra}{\it Chandra\rm}

\newcommand\nodata{ ~$\cdots$~ }

\long\def\symbolfootnote[#1]#2{\begingroup\def\thefootnote{\fnsymbol{footnote}}\footnote[#1]{#2}\endgroup}

\title[The Gas Fraction of Luminous Clusters]{\emph{Chandra} Measurements of a Complete Sample of X-ray Luminous Galaxy Clusters: the Gas Mass Fraction}
\author[D. Landry et al.]{D. Landry$^{1}$\thanks{Email: d.landry@uah.edu}, M. Bonamente$^{1,2}$, P. Giles$^{3}$, B. Maughan$^{3}$, and M. Joy$^{2}$\\
$^{1}$Physics Department, University of Alabama in Huntsville, Huntsville, AL, USA 35899\\
$^{2}$NASA National Space Science and Technology Center, Huntsville, AL, USA 35805\\
$^{3}$HH Wills Physical Laboratory, University of Bristol, Bristol, UK}

\begin{document}
\date{Accepted . Received ; in original form }

\pagerange{\pageref{firstpage}--\pageref{lastpage}} \pubyear{2012}

\maketitle

\label{firstpage}

%

\begin{abstract}
We present \chandra\ X-ray measurements of the gas mass fraction out to $r_{500}$ for a complete sample
of the 35 most luminous clusters from the Brightest Cluster Sample and the Extended Brightest Cluster 
Sample at redshift $z=0.15-0.30$. The sample includes relaxed and unrelaxed clusters, and the data were analysed
independently using two pipelines and two different models for the gas density
and temperature. We measure an average of
$f_{\textrm{gas}}(r_{500}) = 0.163\pm{0.032}$, which is in agreement with the cosmic baryon fraction 
($\Omega_{\textrm{b}} / \Omega_{\textrm{M}} = 0.167\pm{0.006}$) at the  1$\sigma$ level, after adding the stellar baryon fraction.
Earlier studies 
reported gas mass fractions significantly 
lower than the cosmic baryon fraction at $r_{500}$,
and in some cases higher values that are consistent with the
cosmic baryon fraction towards the virial radius.
In this paper we show that the most X-ray luminous clusters
in the redshift range $z=0.15-0.30$ have a gas mass fraction that is consistent with the
cosmic value at $r_{500}$.
\end{abstract}

\section{Introduction}
Clusters of galaxies are the largest known bound systems in the Universe and are formed
from the collapse of primordial density fluctuations \citep{press1974,white1978}.
The intracluster medium (ICM) is in the form of a hot ionized gas at $\sim10^{8}$ K, and 
it contains most of the baryons in clusters. 
The remaining baryons are in stars and  
intracluster light, and  account for a few percent of the total mass \citep{lin2003,gonzalez2007,giodini2009}.
The total cluster baryon fraction is therefore a combination of the baryons in stars 
and ICM, i.e., $f_{\textrm{b}} = f_{\textrm{stars}} + f_{\textrm{gas}}$.
Since clusters are extremely large and massive, baryons and dark matter originated from
approximately the same comoving volume, and thus it is believed that their ratio should be representative
of the Universe \citep[e.g.,][]{metzler1994}. 
A number of studies have used the baryonic mass fraction to measure $\Omega_{\textrm{M}}$ and 
found evidence for a low density Universe
\citep{fabian1991,white1991, white1993, briel1992,david1995,white1995,evrard1997,pen1997,ettori1999,mohr1999,grego2001,allen2002,allen2004,laroque2006,allen2008,ettori2009}.
Measuring the evolution of the gas mass fraction over a large redshift range, $f_{\textrm{gas}}(z)$, 
allows cosmological quantities to be constrained
(for recent reviews see \citealt{allen2008,ettori2009} and references therein).

Current studies indicate that
the cluster baryon fraction is typically lower than the cosmic baryon fraction 
as measured by the $\Omega_{\textrm{b}} / \Omega_{\textrm{M}}$ parameter
\citep{vikhlinin2006,afshordi2007,arnaud2007,giodini2009,sun2009,umetsu2009,rasheed2010,komatsu2011}.
This observation has raised questions on the whereabouts of  these \emph{missing baryons} \citep{rasheed2010}. 
A possible explanation is that Active Galactic Nuclei (AGN) feedback may push the ICM towards 
the cluster outskirts
\citep{metzler1994,takizawa1998,bialek2001,valdarnini2003,mccarthy2007,cavaliere2008,bode2009}.

To address this issue we measured the gas mass fraction
for a complete sample of massive clusters at $z=0.15-0.30$ from the  
\emph{Brightest Cluster Sample} and its extension \citep{ebeling1998,ebeling2000,dahle2006}. 
The background of the surface brightness and temperature profiles 
limits the radius out to which  masses can be measured accurately.
Therefore we chose to limit our measurements for the entire sample to $r_{500}$, 
the radius within which the mass density
is $500$ times the critical density of the universe at the cluster's redshift. 
In cases of long exposures and high quality
data, it is possible to use \chandra\
to detect clusters beyond $r_{500}$. For example,  in \cite{bonamente2012} we report the
detection of A1835, one of the clusters in this sample, out to the virial radius.
We calculate the total cluster baryon fraction by adding the baryons from stars and galaxies, and find that
the baryon content in these high-luminosity clusters is consistent with the cosmic ratio
$\Omega_{\textrm{b}} / \Omega_{\textrm{M}}$ at $r_{500}$.

This paper is structured in the following way: Section \ref{sec:sample_of_clusters} briefly describes the
sample of clusters, Section \ref{sec:reduction} discusses the \chandra\ data reduction, and Section \ref{sec:models}
reviews the modelling of the X-ray data. We present our results in Section \ref{sec:results}, possible
systematic effects in Section \ref{sec:systematics}, a comparison with
other studies in Section~\ref{sec:comparison}, and discussion
and conclusions in Section \ref{sec:conclusion}. 
Throughout this paper we assume a cosmology based on WMAP7 results with $H_{0}=70.2$ km s$^{-1}$ Mpc$^{-1}$, 
$\Omega_{\textrm{M}}=0.27$, and $\Omega_{\Lambda}=0.73$ \citep{komatsu2011}.

\section{Cluster Sample}
\label{sec:sample_of_clusters}
The sample we chose for this analysis
is composed of  35 clusters from the \cite{dahle2006} sample (Table \ref{tab:sample}),
a 90\% complete sample of clusters with X-ray luminosities in the $0.1-2.4$ keV band of
$L_{\textrm{X,\,keV}}$ $\ge 6 \times 10^{44}$~erg s$^{-1}$ (for a concordance $\Lambda$CDM 
universe with $h=0.7$)
from the Brightest Cluster Sample \cite[BCS,][]{ebeling1998} 
and the Extended BCS \cite[eBCS,][]{ebeling2000} in the $z=0.15-0.30$ redshift range.
These clusters are estimated to have masses of $M_{180} \geq 5\times10^{14}$ $h^{-1}$ $\textrm{M}_{\odot}$, making
them the most massive in the (e)BCS sample \citep{dahle2006}.
All of the clusters from this sample have archival \chandra\ data available 
for our analysis.
The majority of the clusters have been observed with the ACIS imaging array (ACIS-I), and 6 with the 
ACIS spectroscopic array (ACIS-S). 
With this sample of clusters, we aim to measure the gas mass fraction out to $r_{500}$ using high 
S/N X-ray data from \chandra, and then calculate the total baryon fraction using the stellar fraction
from \cite{gonzalez2007} and \cite{giodini2009}.

\begin{table*}\footnotesize
\caption{Cluster Sample}
\begin{tabular}{lcccccc}
\hline
\hline
\multirow{2}{*}{Cluster} & \multirow{2}{*}{$z$} & $D_{\textrm{A}}$ & $N_{\textrm{H}}$ & \multirow{2}{*}{obsID} & Exposure & \multirow{2}{*}{Dynamical State}\\
        &     & (Mpc)   & ($10^{20}$ cm$^{-2}$) & & (ks) &\\
\hline
A115		& 0.1971 	& 673.9 	& 5.36 & 3233		& 43.6			& Unrelaxed\\ 
A1423		& 0.2130 	& 716.1 	& 1.81 & 538 		& 35.1			& Unrelaxed\\ 
A1576  		& 0.2790 	& 876.1 	& 1.08 & 7938 		& 15.0			& Unrelaxed\\ 
A1682  		& 0.2260 	& 749.5 	& 1.04 & 11725 		& 19.6			& Unrelaxed\\ 
A1758        	& 0.2790 	& 876.1 	& 1.03 & 7710 		& 7.0			& Unrelaxed\\ 
A1763        	& 0.2230 	& 741.9 	& 0.82 & 3591 		& 17.0			& Unrelaxed\\ 
A1835        	& 0.2532 	& 816.3 	& 2.04 & 6880 6881 7370 & 167.1			& Relaxed\\ 
A1914        	& 0.1712 	& 601.8 	& 1.06 & 3593 542 	& 16.6			& Unrelaxed\\ 
A2111        	& 0.2290 	& 757.1 	& 1.84 & 11726 544 	& 30.9			& Unrelaxed\\ 
A2204        	& 0.1520 	& 545.5 	& 5.67 & 7940 		& 70.5			& Relaxed\\ 
A2219        	& 0.2256 	& 748.5 	& 1.76 & 896 		& 41.5			& Unrelaxed\\ 
A2261        	& 0.2240 	& 744.4 	& 3.19 & 5007 550 	& 22.5			& Relaxed\\ 
A2390        	& 0.2329 	& 766.8 	& 6.21 & 4193 		& 19.7			& Relaxed\\ 
A2552        	& 0.3017 	& 925.9 	& 4.60 & 11730 3288 	& 22.7			& Unrelaxed\\ 
A2631        	& 0.2780 	& 873.8 	& 3.55 & 11728 3248 	& 25.0			& Unrelaxed\\ 
A267         	& 0.2310 	& 762.1 	& 2.75 & 1448 		& 7.4			& Unrelaxed\\ 
A520         	& 0.1990 	& 679.0 	& 5.65 & 528 4215 9424 9426 9430 & 368.0	& Unrelaxed\\ 
A586         	& 0.1710 	& 601.2 	& 4.89 & 11723 530 	& 15.8			& Unrelaxed\\ 
A611         	& 0.2880 	& 896.1 	& 4.46 & 3194 		& 15.4			& Relaxed\\ 
A665         	& 0.1819 	& 632.1 	& 4.32 & 3586 		& 14.3			& Unrelaxed\\ 
A68          	& 0.2546	& 819.7 	& 4.96 & 3250 		& 9.2			& Unrelaxed\\ 
A697         	& 0.2820 	& 882.8 	& 2.93 & 4217 		& 15.4			& Unrelaxed\\ 
A773         	& 0.2170 	& 726.5 	& 1.28 & 3588 5006 533 	& 37.3			& Unrelaxed\\ 
A781         	& 0.2987 	& 919.4 	& 1.65 & 534 		& 9.9			& Unrelaxed\\ 
A963         	& 0.2060 	& 697.7 	& 1.25 & 903 		& 23.0			& Relaxed\\ 
MS~1455+2232    & 0.2578 	& 827.2 	& 3.18 & 4192 543 	& 79.2			& Relaxed\\ 
RX~J0437.1+0043 & 0.2850 	& 889.5 	& 5.50 & 11729 7900 	& 41.7			& Relaxed\\ 
RX~J0439.0+0715 & 0.2300 	& 759.6		& 9.18 & 1449 3583 	& 23.7			& Relaxed\\ 
RX~J1720.1+2638 & 0.1640 	& 581.0 	& 3.36 & 3224 4361 	& 41.7			& Relaxed\\ 
RX~J2129.6+0005 & 0.2350 	& 772.1 	& 3.63 & 552 9370 	& 35.9			& Relaxed\\ 
Z2089        	& 0.2347 	& 771.3 	& 2.86 & 10463 7897 	& 39.4			& Relaxed\\
Z3146        	& 0.2906 	& 901.8 	& 2.46 & 909 9371	& 81.3			& Relaxed\\ 
Z5247        	& 0.2300 	& 759.6 	& 1.61 & 11727 539 	& 19.2			& Unrelaxed\\
Z5768        	& 0.2660 	& 846.4 	& 1.49 & 7898 		& 10.4			& Unrelaxed\\
Z7215        	& 0.2897 	& 899.9 	& 1.40 & 7899 		& 13.0			& Unrelaxed\\ 

\hline
\end{tabular}
\label{tab:sample}
\end{table*}

\section{\textit{Chandra} Observations and Data Reduction}
\label{sec:reduction}
Data reduction was done using the Chandra
Interactive Analysis of Observations (\textsc{ciao}\footnote{http://cxc.harvard.edu/ciao/}) 
software version $4.2$ and 
the Chandra Calibration Database (\textsc{caldb}\footnote{http://cxc.harvard.edu/caldb/}) version $4.3.1.$
The data reduction and subsequent analysis was performed by two separate pipelines:
one was developed by D.~Landry and M.~Bonamente, and  makes use of a Markov chain Monte Carlo
method of analysis \cite[e.g.,][]{bonamente2004,bonamente2006}, the other 
was developed by P.~Giles and B.~Maughan \cite[additional details are provided in][]{giles2012}.
Results from the two analyses, including temperature profiles and mass measurements, were found to be in
statistical agreement, thus providing confidence on the results provided in this paper
and in others to follow using the same sample \cite[e.g.,][]{giles2012}.

\begin{figure}
\centering
\includegraphics[angle=-90,width=2.75in]{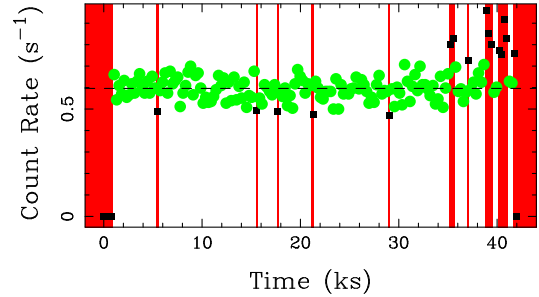}
\caption{A sample lightcurve for observation 9371 of \textit{Zwicky~3146},
showing excluded time intervals due to solar flares (red boxes).
Lightcurves are created over the whole energy range, $0.3-12.0$ keV, using the local background region, $450-900^{\prime\prime}$ in this case.
The mean count rate is $0.596$ s$^{-1}$ (dashed line) and the filtered exposure time is $36.3$ ks.}
\label{fig:lcurve}
\end{figure}

As part of the data reduction, corrections were made for afterglows, charge transfer inefficiency (CTI), 
bad pixels and solar flares. Afterglows are caused from cosmic rays building up charge on the CCD, bad pixels take
into account hot pixels and afterglow events, and CTI is
due to proton damage to the ACIS chips that reduces the
energy resolution. The variability of solar activity can cause 
periods of high background which need to be filtered out. A common way of removing these periods
of solar flares is to follow the lightcurve filtering method of \cite{markevitch2003}. Before the lightcurve can
be created, point sources of high and variable emission need to be excluded. The lightcurve is then created
over the energy range $0.3-12.0$ keV for ACIS-I observations and $2.5-6.0$ keV for ACIS-S observations on a selected region of the \chandra\ CCD's used as the
local background. A sample lightcurve is shown in  Figure~\ref{fig:lcurve}.
This lightcurve was filtered with an iterative algorithm (\texttt{deflare} command in \textsc{ciao}) 
that removes time intervals
outside the $3\sigma$ range of the mean.
The solid line shows the mean count rate for the observation and the red boxes show the regions
that were filtered out. Most observations were taken in VFAINT mode, and in this case
we applied VFAINT cleaning to both the cluster and blank-sky observations.

\begin{figure}
\centering
\includegraphics[angle=-90,width=2.75in]{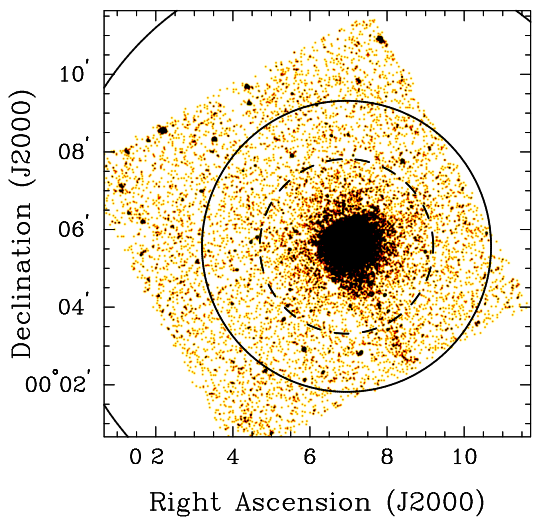}
\caption{The local background region used for ACIS-I observation 9371 of \textit{Zwicky~3146}. The annulus used for the local background is from $450-900^{\prime\prime}$ (solid lines).
The dashed line indicates $r_{500}$ and is approximately $270^{\prime\prime}$.}
\label{fig:localbkgd}
\end{figure}

Images were created in the $0.7-7.0$ keV band to measure
the surface brightness as a function of radius. This energy band was chosen since 
in this energy range
the effective
area is highest and the background rate lowest.
The most crucial aspect of the analysis of diffuse sources such as galaxy clusters
is background subtraction.
For this purpose, we use ACIS blank-sky composite event files and
measurements of the local background taken from source-free regions of the cluster observation
\citep{markevitch2003}.
Since all the clusters in the sample have redshifts in the  $z=0.15-0.30$ range,
the emission does not extend across the entire detector, and
we have  a sufficiently large region to obtain a local background
(see Figure \ref{fig:localbkgd}). 
In nearly every cluster the local background was measured from an outer annulus
beyond $r_{200}$, where the surface brightness was approximately constant, i.e.,
it has reached the background level.
There are a few clusters that only have ACIS-S data available. For these
clusters, the local background was taken from the adjacent ACIS-I chips.
Point sources and extended substructures were detected and removed using \texttt{wavdetect} in \textsc{ciao} and 
known sources were removed manually by visual inspection.

\begin{figure}
\centering
\includegraphics[angle=-90,width=3in]{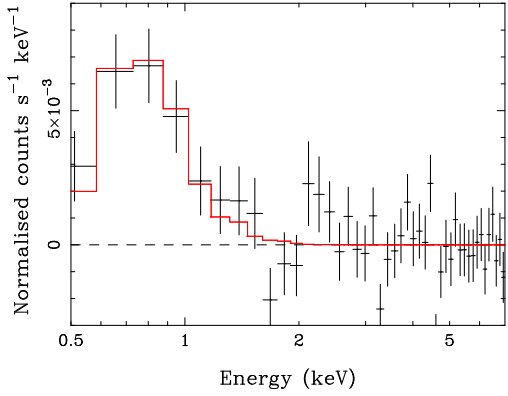}
\caption{Spectrum of the local background (beyond $\sim 750^{\prime\prime}$) for observation 6880 of \textit{Abell~1835}. This shows the soft residuals (below 2 keV) fit with an unabsorbed
thermal plasma model with $kT \sim 0.25$ keV.}
\label{fig:SR}
\end{figure}

The local background of the cluster observation may differ from the ACIS blank-sky
composites, since these observations were done at different times and positions in the sky, and 
the X-ray background has both spatial dependence \cite[e.g.,][]{snowden1997} and is time-variable
\cite[e.g.,][]{takei2008}.
Although the background flux may vary with time, \cite{hickox2006} 
have shown that the ratio of the flux within the $2-7$ keV and
$9.5-12$ keV bands is constant in time.
We can therefore rescale the blank-sky spectrum by the ratio of the count rates in the
$9.5-12$ keV band of cluster and blank-sky observations, 
and obtain a clean
subtraction of the background in the spectral region of interest, which is $0.7-7.0$
keV \citep{hickox2006}.
After subtracting the blank-sky background, residuals may still be present in the soft $0.7-2$ keV band.
These soft residuals may be due to Galactic and extragalactic emission, as well as residual solar flares
that were not removed by lightcurve filtering \citep[e.g.,][]{snowden1997}.
When present, the soft residual spectrum is then fit with a power law and a plasma emission
model, and this model of the soft residuals is taken into account in the cluster
spectra \citep[as done in, e.g.,][]{bulbul2010,hasler2012}.
Figure \ref{fig:SR} shows an example of soft residuals below $2$ keV for observation 6880 of \textit{Abell~1835}. 
Multiple observations were reduced individually to apply the correct calibration to
each dataset.
The cluster surface brightness
profile is obtained from merged images, and the temperature profile from fitting
spectra from different observations simultaneously. 

\section{Modelling and Analysis of the \textit{Chandra} Data}
\label{sec:models}
The total mass of a cluster can be
inferred from the density and temperature of the X-ray emitting ionized plasma, assuming that the hot gas is
in hydrostatic equilibrium with the gravitational cluster potential.
Thus, we can write the total mass as 
\begin{equation}
M_{\textrm{tot}}(r) = \frac{-k_{\textrm{B}}T(r)\,r}{\mu m_{\textrm{p}} G} \left( \frac{d\,\textrm{ln}\,n_{\textrm{e}}(r)}{d\,\textrm{ln}\,r} + \frac{d\,\textrm{ln}\,T(r)}{d\,\textrm{ln}\,r}  \right),
\end{equation}
where $k_{\textrm{B}}$ is the Boltzmann constant, $T(r)$ is the temperature profile, $\mu$ is the mean molecular weight, 
$m_{\textrm{p}}$ is the mass of a proton, $G$ is the gravitational constant, and $n_{\textrm{e}}(r)$ is the number density
of electrons.
The mass of the hot gas is calculated as
\begin{equation}
M_{\textrm{gas}}(r) = 4 \pi \mu_{\textrm{e}} m_{\textrm{p}} \int n_{\textrm{e}}(r)\,r^2\;dr,
\end{equation}
where $\mu_{\textrm{e}}$ is the mean molecular weight of the electrons. 
We use high resolution X-ray imaging and spectroscopy from the \chandra\ X-ray Observatory to
determine $n_{\textrm{e}}(r)$ and $T(r)$. Specifically, the observed X-ray surface brightness is related
to the electron number density by the following equation
\begin{equation}
S_{\textrm{X}}=\frac{1}{4 \pi (1 + z)^3} \int n^{2}_{\textrm{e}}(r) \Lambda_{\textrm{ee}}(T_{\textrm{e}}) \; dl 
\end{equation}
where $z$ is the redshift,
$\Lambda_{\textrm{ee}}(T_{\textrm{e}})$ is the plasma emissivity (counts cm$^3$ s$^{-1}$)
and the integral is along the line of sight through
the cluster.
The linear distance $r$ in these equations is given by $r=\theta D_{\textrm{A}}$, where $\theta$ is the apparent
angular size and $D_{\textrm{A}}$ is the angular diameter
\citep{carroll1992}.
To measure the density from the surface brightness, the angular diameter must be known.
The value of $D_{\textrm{A}}$ can be calculated from the following
equation (valid for $\Omega_{k}=0$):
\begin{equation*}
D_{\textrm{A}}=\frac{c}{(1 + z) H_{0}} \int^z_0 \frac{1}{E(z)} \; dz,
\end{equation*}
where 
$E^2(z)=\Omega_{\textrm{M}}(1+z)^3 + \Omega_{\Lambda}$
and $H_{0}$ is the Hubble constant. The temperature profile is determined from radially-averaged X-ray spectra
fit to an \textsc{apec} optically-thin emission model  \citep{smith2001}, with Galactic \textsc{h i} column
density measured from the Leiden-Argentine-Bonn survey \citep{kalberla2005}. We used the
\textsc{xspec} software version 12.6.0s \citep{arnaud1996}. 

We use the \cite{vikhlinin2006} model that describes the density and temperature of the cluster to 
fit the \chandra\ data.
The three-dimensional gas density is modelled as a generalisation of the $\beta$-model,
\begin{equation}
\label{eq:vikhdensity}
\begin{aligned}
n_{\textrm{p}}n_{\textrm{e}} = & \;n_{0}^{2} \frac{(r/r_{\textrm{c}})^{-\alpha}}{(1+r^{2}/r_{\textrm{c}}^{2})^{3\beta-\alpha/2}} \frac{1}{(1+r^{\gamma}/r_{\textrm{s}}^{\gamma})^{\varepsilon/\gamma}} +\\
   &    \;\frac{n_{02}^{2}}{(1+r^{2}/r_{\textrm{c}}^{2})^{3\beta_{2}}}
\end{aligned}
\end{equation}
and uses a total of ten parameters. The model used for the temperature profile is given by
the phenomenological function
\begin{equation}
\label{eq:vikhtemp}
\begin{aligned}
T(r) = & \;T_{0} \frac{(r/r_{\textrm{cool}})^{a_{\textrm{cool}}}+(T_{\textrm{min}}/T_{0})}{(r/r_{\textrm{cool}})^{a_{\textrm{cool}}}+1} \times\\
       & \;\frac{(r/r_{\textrm{t}})^{-a}}{[1+(r/r_{\textrm{t}})^{b}]^{c/b}},
\end{aligned}
\end{equation}
which has eight parameters and thus enough degrees of freedom
to model nearly any smooth temperature distribution. The
second term in the temperature profile describes the region outside of the cool-core as a
broken power law with a transition region. Therefore, the \cite{vikhlinin2006} model has a
total of 18 parameters. In our analysis, we used the following constraints for all clusters:
$n_{02}=0$, $\alpha=0$, $\gamma=3$, and $\varepsilon < 5$. For certain clusters
with lower S/N, we  fixed additional parameters.

To assess biases from the use of a particular parameterization of the thermodynamic quantities,
we also model all clusters with the \cite{bulbul2010} model. 
This model gives analytic functions for temperature,
density, and gas pressure, assuming a polytropic equation of state and
hydrostatic equilibrium outside of the cluster core. Similar to the
\cite{vikhlinin2006} model, it accounts for cooling of the gas in the core, and the temperature
profile is
\begin{equation}
\label{eq:polytemp}
T(r)=T_{0} \bigg( \frac{1}{(\beta-2)}\frac{(1+r/r_{\textrm{s}})^{\beta-2}-1}{r/r_{\textrm{s}}(1+r/r_{\textrm{s}})^{\beta-2}} \bigg) \;\tau_{\textrm{cool}}.
\end{equation}
The function
\begin{equation}
\label{eq:tempcool}
\tau_{\textrm{cool}}=\frac{(r/r_{\rm cool})^{a_{\textrm{cool}}} + \xi}{(r/r_{\rm cool})^{a_{\textrm{cool}}} + 1}
\end{equation}
has been
adopted from \cite{vikhlinin2006}. 
In Equation \ref{eq:tempcool}, the quantity $\xi=T_{\textrm{min}}/T_{0}$ is a free parameter
($0$~\textless~$\xi$~\textless~$1$) that measures the amount of central cooling.
With this temperature profile, the gas
density profile can be determined from the polytropic equation of state,
\begin{equation}
\label{eq:polydensity}
n_{\textrm{e}}(r)=n_{\textrm{e0}} \bigg( \frac{1}{(\beta-2)} \frac{(1+r/r_{\textrm{s}})^{\beta-2} - 1}{r/r_{\textrm{s}}(1+r/r_{\textrm{s}})^{\beta-2}} \bigg)^n \tau_{\textrm{cool}}^{-1},
\end{equation}
The \cite{bulbul2010} model therefore has a total of $8$ parameters.
The comparison of results from the two models can be found in Section \ref{sec:systematics}.

The emissivity for thermal bremsstrahlung emission at energy $E$ is 
\begin{equation*}
\Lambda_{ee}  \propto n^{2} T^{-1/2} e^{-E/k_{\textrm{B}}T},
\end{equation*} 
where $n$ is the number density and $T$ is the temperature.
 The two X-ray observables, $S_{\textrm{X}}$ and $T$, are therefore not completely independent. 
To determine the best-fit parameters of the models, we use a Markov chain Monte Carlo (MCMC) 
method with
a Metropolis-Hastings selection algorithm described in \cite{bonamente2004}.
Although both models have parameters that can be correlated to one another \citep[e.g.,][]{hasler2012},
the MCMC method  accounts for these correlations, and model parameters and all derivative
quantities such as masses and the gas mass fraction can be computed accurately.

In the analysis of the surface brightness and temperature profile data 
we follow \cite{bulbul2010} and \cite{hasler2012} and add
a 1\% systematic uncertainty for 
each bin of the surface brightness and a 10\% systematic uncertainty for each
bin of the temperature data.
In 
 Section \ref{sec:systematics}  we describe additional sources of systematic
errors that can affect our measurements.
Selected temperature fits
for both the \cite{vikhlinin2006} and \cite{bulbul2010} model are shown in Figure
\ref{fig:T-fits-few} (see Appendix \ref{sec:appendixA} for temperature
profiles and fits for all clusters).
To measure masses at $r_{500}$, we need to fit the data beyond this
radius to constrain the slope of the temperature profile. 
Out of the entire sample of 35 clusters, 
28 have temperature and surface brigthness measurements out to or beyond
$r_{500}$, and 
only 7 required slight extrapolation out to 
$r_{500}$ (see Appendix \ref{sec:appendixA} for additional details).

\begin{figure*}
\centering
\includegraphics[angle=-90,width=2.85in]{./FIG/a115_poly_vikh_temp.ps}
\includegraphics[angle=-90,width=2.85in]{./FIG/a1914_poly_vikh_temp.ps}
\includegraphics[angle=-90,width=2.85in]{./FIG/a2204_poly_vikh_temp.ps}
\includegraphics[angle=-90,width=2.85in]{./FIG/a963_poly_vikh_temp.ps}
\caption{Temperature profiles for selected clusters using the \protect\cite{vikhlinin2006} model (blue)
and the \protect\cite{bulbul2010} model (red). The solid lines show the best-fit values, and
the hatched region is the 68.3\% confidence interval. 
}
\label{fig:T-fits-few}
\end{figure*}

\section{Measurement of the Gas Mass Fraction}
\label{sec:results}
\subsection{Results for the Entire Sample}
\begin{figure}
\centering
\includegraphics[angle=-90,width=3in]{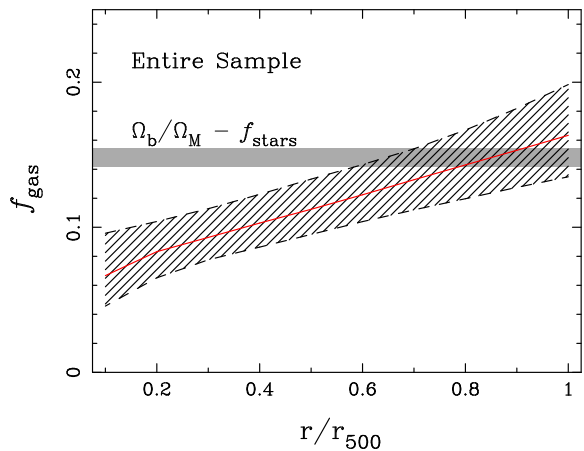}
\caption{Average gas mass fraction profile for the \protect\cite{vikhlinin2006} model for
all the clusters in the sample,
 with $f_{\textrm{gas}}(r_{500})=0.163\pm{0.032}$. The red line is the median and the hatched region
is the 68.3\% confidence interval from the combined Monte Carlo
Markov chains for all clusters. The grey envelope is the difference of the cosmic baryon fraction
and the fraction of baryons in stars and galaxies.
Using the results from \protect\cite{komatsu2011} and \protect\cite{giodini2009},
we use the value of $\Omega_{\textrm{b}} / \Omega_{\textrm{M}}-f_{\textrm{stars}}=0.148\pm{0.006}$.}
\label{fig:vikh-fgas-profile-sample}
\end{figure}

The gas mass, total mass, and gas mass fraction for each cluster are reported
in Table \ref{tab:masses}.
We calculated the median and 68.3\% confidence interval for the average
gas mass fraction from the combined chains of all clusters, to find
\begin{equation}
\begin{cases}
f_{\textrm{gas}}(r_{2500})=0.110\pm{0.017}\\
f_{\textrm{gas}}(r_{500})=0.163\pm{0.032}.\
\end{cases}
\end{equation}
The average radial profile of the gas mass fraction for all clusters is shown in 
Figure \ref{fig:vikh-fgas-profile-sample} and shows an increase with radius as also found 
in previous studies, e.g., 
\cite{vikhlinin2006} and \cite{rasheed2010}.

Since clusters are the largest gravitationally bound structures in the Universe,
the baryon fraction should be representative of the cosmic ratio $\Omega_{\textrm{b}} / \Omega_{\textrm{M}}$.
Current measurements from WMAP indicate that the cosmic baryonic
fraction is
\begin{equation}
\Omega_{\textrm{b}} / \Omega_{\textrm{M}}=0.167\pm{0.006}
\end{equation}
\citep{komatsu2011}.
To accurately
compare the gas mass fraction with the cosmic baryon fraction,
 baryons in stars and galaxies need to be taken into account.
\cite{giodini2009} measured the baryon fraction of stars and
galaxies and determined this stellar mass fraction
as a function of $M_{500}$, finding
\begin{equation}
f_{\textrm{stars},\,500}=0.019\pm{0.002}
\end{equation}
for clusters with mass $M(r_{500}) \simeq 7.1\times 10^{14}$ $\textrm{M}_{\odot}$.
This result is slightly higher than the value
reported by \cite{gonzalez2007}, who measure $f_{\textrm{stars},\,500}\simeq 0.012$ for
the same mass range.
Including the \cite{giodini2009} results in our measurements of the gas mass fraction,
we estimate the baryon fraction $f_{\textrm{b}}(r_{500})$ for our sample as
\begin{equation}
f_{\textrm{b}}(r_{500}) = 0.182 \pm 0.032.
\end{equation}
The difference between the average baryon fraction measured at $r_{500}$
and the cosmic baryon fraction from WMAP is therefore
\begin{equation*}
f_{\textrm{b}}(r_{500})-\Omega_{\textrm{b}} / \Omega_{\textrm{M}}=+0.015\pm{0.033},
\end{equation*}
i.e., our measurements match the known
cosmic baryon fraction at the 1$\sigma$ level at $r_{500}$, 
as also shown in Figure \ref{fig:vikh-fgas-profile-sample}. 
The measurement of the baryon fraction using the
\cite{bulbul2010} model also agrees with $\Omega_{\textrm{b}} / \Omega_{\textrm{M}}$
(see Section~\ref{sec:systematics} for discussion).

\subsection{Results for Relaxed and Unrelaxed Clusters}
\label{sec:relaxed_unrelaxed}
We used three parameters to determine if a cluster can be classified as dynamically relaxed 
and host a cool-core: centroid shift for the dynamical state, central cooling time and cuspiness 
for the presence of a cool-core. 
For clusters with high quality data, the central
cooling time is the best method for determining if it has a cool-core or not \citep{hudson2010}.
We calculated the central cooling time as
\begin{equation}
t_{\textrm{cool}} \; = \; 8.5 \times10^{10} \; \textrm{yr} \;\, \bigg( \frac{n_{\textrm{p}}}{10^{-3} \; \textrm{cm}^{-3}} \bigg)^{-1} \bigg( \frac{T_{\textrm{g}}}{10^{8} \; \textrm{K}}\bigg)^{1/2},
\label{eq:tcool}
\end{equation}
where $n_{\textrm{p}}$ is the number density of protons and $T_{\textrm{g}}$ is the central temperature of the gas
\citep{sarazin1988}, using the best-fitting temperature profile at $r=0.048\;r_{500}$ as the central temperature.
Cuspiness is defined as the logarithmic derivative of the density profile,
\begin{equation}
\alpha \, = \, \frac{d \, \textrm{log} \, n_{\textrm{e}}}{d \, \textrm{log} \, r},
\label{eq:cuspiness}
\end{equation}
evaluated at $r=0.04\,r_{500}$ \citep{vikhlinin2007}.
\cite{hudson2010} indicate that, for low quality data, this parameter is a good indicator for the
presence of a cool-core.
The centroid shift is defined as the
standard deviation of the distance between the peak X-ray emission and the centroid
\citep{poole2006}. We measured the centroid shifts in annular bins centered on the
X-ray peak decreasing by $0.05\,r_{500}$, as suggested by \cite{poole2006}. We calculated the central
cooling times, cuspiness, and centroid shifts for all the clusters in the sample, and we
classified relaxed clusters as those which satisfied the following three conditions:
(i) $\langle w\rangle < 0.009\,r_{500}$, (ii) $\alpha > 0.65$, and  (iii) $t_{\textrm{cool}} < 6.5$~Gyr
(see \citealt{giles2012} for more details). 
This classification results in 13 relaxed clusters and 22 unrelaxed clusters, as shown in 
Table \ref{tab:sample}.

\begin{figure*}
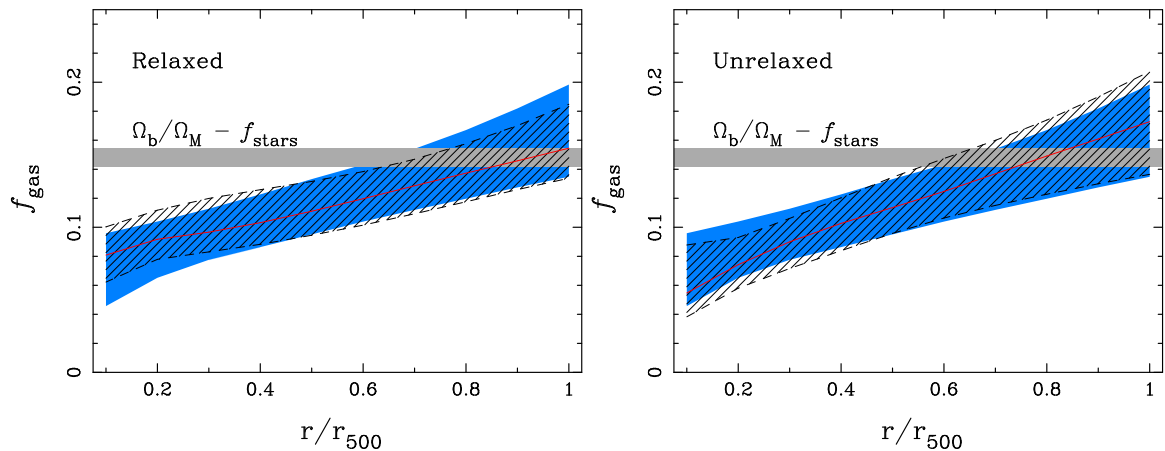

\centering
\includegraphics[angle=-90,width=3in]{./FIG/fgas_profile_WMAP_vikh_relaxed_subsample.ps}
\includegraphics[angle=-90,width=3in]{./FIG/fgas_profile_WMAP_vikh_unrelaxed_subsample.ps}
\caption{Left: Average gas mass fraction profile using the \protect\cite{vikhlinin2006} model for
the relaxed clusters, with $f_{\textrm{gas}}(r_{500})=0.155\pm{0.026}$. 
Right: Average gas mass fraction profile for the unrelaxed clusters,
with $f_{\textrm{gas}}(r_{500})=0.173\pm{0.036}$. 
The red line is the median and the hatched region
is the 68.3\% confidence interval. 
The blue region is the average $f_{\textrm{gas}}$ profile for all clusters.
The grey envelope is the difference of the cosmic baryon fraction
and the fraction of baryons in stars and galaxies,
$\Omega_{\textrm{b}} / \Omega_{\textrm{M}}-f_{\textrm{stars}}=0.148\pm{0.006}$ \protect\citep{giodini2009,komatsu2011}.}
\label{fig:vikh-fgas-profile-subsamples}
\end{figure*}

We calculated the gas mass fraction at 
$r_{2500}$ and $r_{500}$ for both subsamples of relaxed clusters and unrelaxed clusters
(see Table \ref{tab:results}). 
Our results show that at both radii the gas mass fractions for relaxed clusters are in excellent 
agreement with that of unrelaxed clusters, and that both sub-samples
are  statistically consistent with the cosmic
baryon fraction at $r_{500}$ (see Figure \ref{fig:vikh-fgas-profile-subsamples}).
The agreement of the gas mass fraction between relaxed and unrelaxed clusters may be an 
indication that hydrostatic mass estimates are reliable for unrelaxed clusters
out to $r_{500}$.  \cite{giles2012} discusses further the reliability of hydrostatic mass estimates.

\begin{table}
\centering
\caption{Measurements of $f_{\textrm{gas}}$ using the \protect\cite{vikhlinin2006} Model}
\begin{tabular}{lcc}
\hline
\hline
\multirow{2}{*}{Sample} & \multicolumn{2}{c}{$f_{\textrm{gas}}$}\\
       & $r_{2500}$ & $r_{500}$\\
\hline
Relaxed      & $0.111\pm{0.017}$ & $0.155\pm{0.026}$\\
Unrelaxed    & $0.108\pm{0.017}$ & $0.173\pm{0.036}$\\
\hline
All Clusters & $0.110\pm{0.017}$ & $0.163\pm{0.032}$\\
\hline
\end{tabular}
\label{tab:results}
\end{table}

\section{Systematics and Physical Processes Affecting the Gas Mass Fraction}
\label{sec:systematics}
This section describes certain sources of systematic error and physical processes that may affect X-ray measurements
of the gas mass fraction.

\subsection{Assumption of hydrostatic equilibrium}
One way in which the gas mass fraction can be affected is from the assumption of hydrostatic equilibrium.
If there is an additional component of pressure that is non-thermal, then
this must be taken into account when applying the equation of hydrostatic equilibrium.
The hydrostatic equilibrium equation is used to determine the total mass of the
cluster from X-ray observations,
\begin{equation}
\label{eq:dPdr}
M_{\textrm{tot}}(r)=-\frac{r^{2}}{\rho G} \frac{dP}{dr},
\end{equation}
where $\rho$ and $P$ are the density and pressure, respectively, of the hot gas.
Continued accretion of the gas onto clusters along filaments, mergers, and supersonic motions of galaxies
through the intracluster medium are believed to cause gas motions which give rise to non-thermal pressure
\citep{lau2009}.
Suppose that the pressure of the hot gas consists of a thermal and a non-thermal component,
$P=P_{\textrm{th}}+P_{\textrm{non-th}}$. Then, the equation for total mass becomes
\begin{equation}
\label{eq:dPnonth}
M_{\textrm{tot}}(r)=-\frac{r^{2}}{\rho G} \bigg( \frac{dP_{\textrm{th}}}{dr} + \frac{dP_{\textrm{non-th}}}{dr} \bigg),
\end{equation}
and yields a higher mass compared to the mass from thermal pressure only. Separating the mass
into components which correspond to different pressure terms, the total mass can be written
as
\begin{equation}
M_{\textrm{tot}}=M_{\textrm{th}} + M_{\textrm{non-th}}.
\end{equation}
\cite{lau2009} shows that the hydrostatic
mass underestimates the true mass for several simulated clusters, especially at large radii.
Gas motions are expected to cause
a non-thermal pressure in the amount of $\gtrsim$ $5\%-15$\% of the thermal pressure,
and this non-thermal pressure will cause an underestimate of mass at large radii of $8\%\pm{2}\%$ for relaxed
systems and $11\%\pm{6}\%$ for unrelaxed systems at $r_{500}$ \citep{lau2009}.
In the presence of non-thermal pressure the true mass of the cluster is given by 
Equation~\ref{eq:dPnonth}, and the
use of Equation~\ref{eq:dPdr} leads to an underestimate of the mass, and therefore an \emph{overestimate} of the
gas mass fraction.
\cite{giles2012} compared X-ray hydrostatic masses with weak-lensing masses for these clusters and found that the total mass
obtained through X-ray measurements is underestimated by a factor of $1.21\pm{0.23}$ and $1.41\pm{0.15}$ 
for relaxed and unrelaxed clusters, respectively. 
This comparison would indicate a departure from hydrostatic equilibrium that causes the gas mass fraction to be 
overestimated.
Based on the \cite{lau2009} results, we assess a systematic uncertainty of $+10\%$ on $M_{\textrm{tot}}(r_{500})$ for
all clusters. This results in a possible systematic error of $-10\%$ on $f_{\textrm{gas}}(r_{500})$. 

\subsection{Uncertainties in the calibration of the Chandra data}
Uncertainties in the calibration of the \chandra\ X-ray data
 can also affect the measurement of the gas mass fraction.
The efficiency of the ACIS detector has a spatial dependency,
and deviates from being uniform at the $\pm{1\%}$ level \citep{bulbul2010}. Therefore, we 
added a $\pm{1\%}$ uncertainty
to each data point in the surface brightness data prior to the analysis. 

The temperature profile used in the analysis is also subject to various sources of systematic uncertainty. 
One source of uncertainty is from the subtraction of the local background. 
To subtract the background, we first have to rescale the blank-sky spectrum to match that of the
high-energy ($9.5-12$ keV) flux of the cluster observation, as described
in  Section \ref{sec:reduction}.
We estimated the effect of the background subtraction 
using the longest observation
of \textit{Abell~1835} (ObsID 6880) which has the highest S/N. To obtain a clean background subtraction, 
we found that the fractional
correction to the blank-sky spectrum is $-0.04\pm{0.01}$. We measured the temperature in an
outer region out to $\sim$ $r_{500}$ (radii $240-330^{\prime\prime}$) 
using the best-fit correction factor ($-0.04$)
and obtained $kT=5.77$~keV. We then changed the correction to the blank-sky spectrum to $-0.03$ and
$-0.05$ to over- and under-subtract the background by $\pm 1 \sigma$. 
With these values, we found the temperature changed
to $kT=5.38$~keV and $6.23$~keV, respectively. Therefore, we concluded that the uncertainty
in temperature due to the background subtraction is approximately $\pm 7.5$\%.
Another source of error is caused by contamination on the Optical Blocking Filter, which is known to affect 
cluster temperatures by up to 5\% \citep{bulbul2010,hasler2012}.
Adding these errors in quadrature, we use a $\pm{10\%}$ systematic uncertainty in fitting the temperature data.
This additional error was therefore added to each temperature datapoint
prior to the analysis.

\subsection{Uncertainties due to asphericity of the cluster emission}
\begin{figure}
\centering
\includegraphics[angle=-90,width=2.5in]{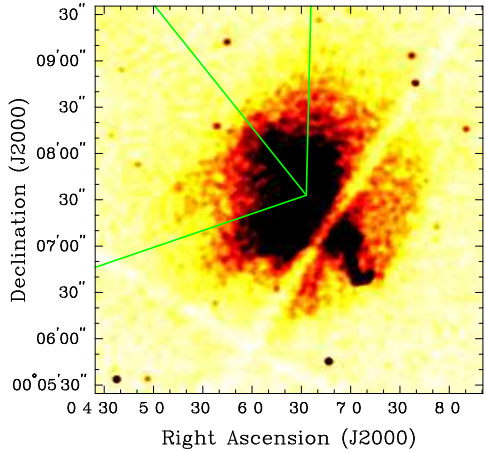}
\caption{Image of \textit{Abell~520} showing the two sections analysed. The regions were selected
by using the temperature map from \protect\cite{govoni2004}.}
\label{fig:a520}
\end{figure}

\begin{figure}
\centering
\includegraphics[angle=-90,width=3in]{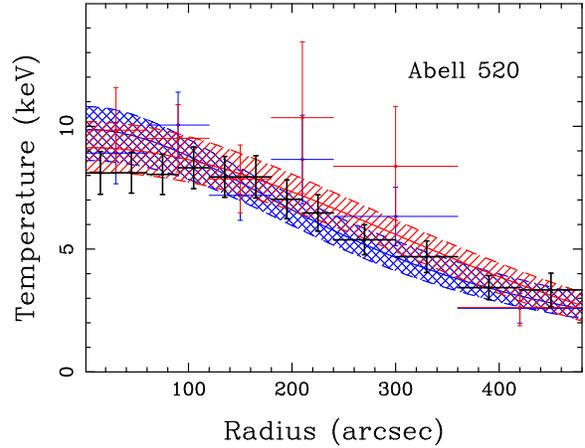}
\caption{Temperature profiles for the two sections analysed from Figure \ref{fig:a520}. The blue
lines correspond to the eastern section and red corresponds to the northern section. The black
data is the temperature profile obtained from analysing the whole cluster.}
\label{fig:T-slices}
\end{figure}

To measure cluster masses, we assume spherical symmetry even though many clusters have 
a disturbed morphology. 
To estimate the uncertainty due to spherical symmetry we considered one of the most
disturbed clusters, \textit{Abell~520}. We used observation ID 9426 since this observation had the best aimpoint for
our task. After performing the \chandra\ data reduction as described in Section \ref{sec:reduction}, this
observation was left with 96.5~ks of filtered exposure time. Using the 2-D temperature map of \textit{Abell~520} from
\cite{govoni2004}, we studied two sections of the cluster (see Figure \ref{fig:a520}): the northern section
(with respect to the azimuthal angle) was chosen because it
 encompasses a gas with temperature $\sim$ $10$ keV, 
while the eastern section was selected because of its cooler temperature of $\lesssim$~8~keV\citep{govoni2004}.
The two sections are representative of an extreme case of 
disturbed dynamical state that causes azimuthal differences in temperature and surface brightness.
We extracted a temperature profile and surface brightness profile for the two sectors,
 and measured masses using only the data within these azimuthal angles. 
Figure \ref{fig:T-slices} shows the temperature fits to the two sections.
For the northern section, we calculated $f_{\textrm{gas}}(r_{2500})=0.109\pm{0.010}$ and 
$f_{\textrm{gas}}(r_{500})=0.179\pm{0.015}$; for the 
eastern section, $f_{\textrm{gas}}(r_{2500})=0.101\pm{0.007}$ and $f_{\textrm{gas}}(r_{500})=0.202\pm{0.015}$ 
(see Table \ref{tab:a520slices} for more information). 
Using these measurements we estimate that there is a $\pm$6\% and $\pm$8\% systematic 
uncertainty in the gas mass fraction due to the assumption
of spherical symmetry at $r_{2500}$ and $r_{500}$, respectively.

\begin{table*}\scriptsize
\centering
\caption{Cluster Properties of Sections of \emph{Abell~520}}
\begin{tabular}{lcccccccccc}
\hline
\hline
& \multicolumn{4}{c}{$\Delta = 2500$} & \multicolumn{4}{c}{$\Delta = 500$}\\
Cluster & $r_{\Delta}$ & $M_{\textrm{gas}}$ & $M_{\textrm{tot}}$ & $f_{\textrm{gas}}$ & $r_{\Delta}$ & $M_{\textrm{gas}}$ & $M_{\textrm{tot}}$ & $f_{\textrm{gas}}$\\
& (arcsec) & (10$^{13} \,\textrm{M}_{\odot}$) & (10$^{14} \,\textrm{M}_{\odot}$) & & (arcsec) & (10$^{13} \,\textrm{M}_{\odot}$) & (10$^{14} \,\textrm{M}_{\odot}$) &\\
\hline
A520
& 166.5 $\pm^{6.0}_{7.2}$ & 3.16 $\pm^{0.22}_{0.25}$ & 2.82 $\pm^{0.31}_{0.35}$ & 0.112 $\pm^{0.006}_{0.005}$ & 367.1 $\pm^{9.3}_{9.2}$ & 10.71 $\pm^{0.30}_{0.29}$ & 6.04 $\pm^{0.47}_{0.44}$ & 0.177 $\pm^{0.009}_{0.009}$\\
Hot Slice
& 168.1 $\pm^{12.3}_{13.1}$ & 3.18 $\pm^{0.44}_{0.44}$ & 2.90 $\pm^{0.68}_{0.63}$ & 0.109 $\pm^{0.011}_{0.009}$ & 375.5 $\pm^{16.4}_{15.3}$ & 11.60 $\pm^{0.59}_{0.57}$ & 6.47 $\pm^{0.88}_{0.76}$ & 0.179 $\pm^{0.015}_{0.015}$\\
Cool Slice
& 165.0 $\pm^{8.7}_{10.9}$ & 2.76 $\pm^{0.29}_{0.34}$ & 2.74 $\pm^{0.46}_{0.51}$ & 0.101 $\pm^{0.008}_{0.006}$ & 338.5 $\pm^{14.5}_{14.9}$ & 9.58 $\pm^{0.56}_{0.58}$ & 4.74 $\pm^{0.63}_{0.60}$ & 0.202 $\pm^{0.016}_{0.014}$\\
\hline
\label{tab:a520slices}
\end{tabular}
\end{table*}

\subsection{Effects of model parameterization}
\begin{figure}
\centering
\includegraphics[angle=-90,width=3in]{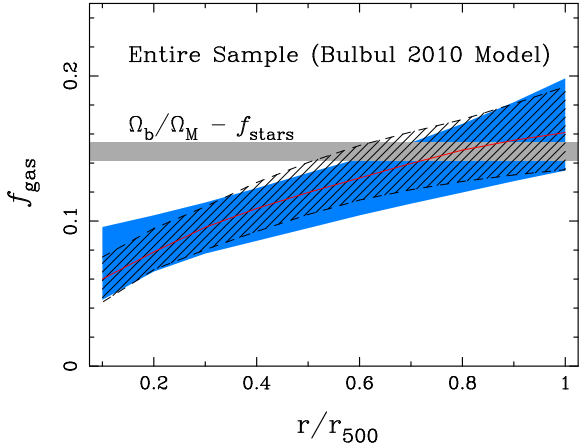}
\caption{Average gas mass fraction profile for the \protect\cite{bulbul2010} model for
all the clusters in the sample with $f_{\textrm{gas}}(r_{500})=0.161\pm{0.029}$. The red line is the median and the hatched region
is the 68.3\% confidence interval. The blue envelope is the sample average $f_{\textrm{gas}}$ using the \protect\cite{vikhlinin2006} model.
The grey envelope is the difference of the cosmic baryon fraction
and the fraction of baryons in stars and galaxies, 
$\Omega_{\textrm{b}} / \Omega_{\textrm{M}}-f_{\textrm{stars}}=0.148\pm{0.006}$
\protect\citep{giodini2009,komatsu2011}.}
\label{fig:poly-fgas-profile-sample}
\end{figure}

Another source of error is due to the choice of model for the X-ray data. 
In addition to the baseline model by \cite{vikhlinin2006},
we repeated all measurements of the gas mass fraction with 
 the \cite{bulbul2010} model 
(see Section \ref{sec:models} for more detail about the models).
Model parameters for both fits 
are reported in Tables \ref{tab:vikh_density}, \ref{tab:vikh_temperature}, and \ref{tab:poly_parameters}
(fixed parameters are shown without error bars), and the mass measurements using both models can be found in 
Table \ref{tab:masses}.
A plot of the average gas mass fraction profile for the entire sample using the \cite{bulbul2010} model is given in
Figure \ref{fig:poly-fgas-profile-sample}, showing an excellent agreement at all
radii with the \cite{vikhlinin2006} model results.

We also compare the average values
of $r_{\Delta}$, $M_{\textrm{gas}}$, $M_{\textrm{tot}}$, and $f_{\textrm{gas}}$ for the two models 
in Table \ref{tab:model_comparison};
plots comparing each measurement can be found in Figure \ref{fig:poly_vs_vikh}.  
By using these two different models to measure
cluster masses we conclude there is a $6\%\pm{2}\%$ uncertainty on $f_{\textrm{gas}}$ at $r_{2500}$ and a $1\%\pm{2}$\% uncertainty
on $f_{\textrm{gas}}$ at $r_{500}$ due to modelling the data, i.e., the two models give the
same answer at large radii.

We also used two independent pipelines to reduce and analyse the same \chandra\ data
and measure masses, one developed by Landry and Bonamente (LB) and one
by Giles and Maughan (GM). The two analyses resulted in the following differences in the measurement
of the gas mass fraction:
\begin{equation*}
\begin{cases}
\begin{aligned}
f_{\textrm{gas}}(r_{2500}): & \frac{\textrm{GM} - \textrm{LB}}{\textrm{GM}} = -2\%\pm{3}\%\\
f_{\textrm{gas}}(r_{500}):  & \frac{\textrm{GM} - \textrm{LB}}{\textrm{GM}} = -3\%\pm{3}\%.
\end{aligned}
\end{cases}
\end{equation*}
This result indicates that our measurements of the gas mass fraction
using the two pipelines
 are consistent with one another. We therefore believe that
our measurements of the gas mass fraction
are robust to the various choices made during the analysis. 

We note that a recent study by \cite{mantz2011} shows how parametric models for the density and temperature
of the ICM introduce an implicit prior due to the assumption of hydrostatic equilibrium. If the models 
are not flexible enough or too general, then the derived scaling relations will be biased towards
self-similarity \citep{mantz2011}. 

\subsection{Uncertainties due to clumping of the gas}
Clumping of the gas will also affect the gas mass fraction. 
The main process for X-ray emission in clusters is through bremsstrahlung emission, which
is proportional to $n^{2}_{\textrm{e}}$. If the gas is clumped, instead of being distributed uniformly, then
the density of electrons $n_{\textrm{e}}$ will be overestimated by making the assumption of a uniform
distribution.
Observations and simulations show that gas clumping may be most evident in the outskirts of clusters
\citep[$r > r_{500}$,][]{mathiesen1999,nagai2011,eckert2012}, though even within $r_{500}$ the gas
mass can be overestimated by $\sim10\%$ due to clumping \citep{mathiesen1999}.
Therefore, our measurements of $f_{\textrm{gas}}$ at $r_{500}$ would remain consistent with the cosmological
value of $\Omega_{\textrm{b}} / \Omega_{\textrm{M}}$ even after accounting 
for a 10\% reduction due to possible clumping of the gas.

\begin{table*}
\centering
\caption{\protect\cite{vikhlinin2006} and \protect\cite{bulbul2010} Model Comparison of Sample Average Cluster Properties}
\begin{tabular}{lcccc}
\hline
\hline
\multirow{2}{*}{Model} & $r_{2500}$ & $M_{\textrm{gas}}$ & $M_{\textrm{tot}}$ & \multirow{2}{*}{$f_{\textrm{gas}}$}\\
      & (arcsec) & ($10^{13} \,\textrm{M}_{\odot}$) & ($10^{14} \,\textrm{M}_{\odot}$) &\\
\hline
\protect\cite{vikhlinin2006} & $137.4\pm{34.7}$ & $2.80\pm{1.50}$ & $2.53\pm{1.38}$ & $0.110\pm{0.017}$\\
\protect\cite{bulbul2010} & $132.4\pm{27.7}$ & $2.73\pm{1.05}$ & $2.36\pm{0.96}$ & $0.115\pm{0.016}$\\
\hline
\hline
\multirow{2}{*}{Model} & $r_{500}$ & $M_{\textrm{gas}}$ & $M_{\textrm{tot}}$ & \multirow{2}{*}{$f_{\textrm{gas}}$}\\
      & (arcsec) & ($10^{13} \,\textrm{M}_{\odot}$) & ($10^{14} \,\textrm{M}_{\odot}$) &\\
\hline
\protect\cite{vikhlinin2006} & $292.2\pm{71.1}$ & $8.07\pm{3.28}$ & $5.05\pm{2.43}$ & $0.163\pm{0.032}$\\
\protect\cite{bulbul2010} & $301.2\pm{59.4}$ & $8.57\pm{2.78}$ & $5.58\pm{2.26}$ & $0.161\pm{0.029}$\\
\hline
\end{tabular}
\label{tab:model_comparison}
\end{table*}

\begin{figure*}
\centering
\includegraphics[angle=-90,width=2.5in]{./FIG/r2500_poly_vikh_relaxed+unrelaxed.ps}
\includegraphics[angle=-90,width=2.5in]{./FIG/r500_poly_vikh_relaxed+unrelaxed.ps}
\includegraphics[angle=-90,width=2.5in]{./FIG/Mgas2500_poly_vikh_relaxed+unrelaxed.ps}
\includegraphics[angle=-90,width=2.5in]{./FIG/Mgas500_poly_vikh_relaxed+unrelaxed.ps}
\includegraphics[angle=-90,width=2.5in]{./FIG/Mtot2500_poly_vikh_relaxed+unrelaxed.ps}
\includegraphics[angle=-90,width=2.5in]{./FIG/Mtot500_poly_vikh_relaxed+unrelaxed.ps}
\includegraphics[angle=-90,width=2.5in]{./FIG/fgas2500_poly_vikh_relaxed+unrelaxed.ps}
\includegraphics[angle=-90,width=2.5in]{./FIG/fgas500_poly_vikh_relaxed+unrelaxed.ps}
\caption{Comparison between measurements using the \protect\cite{vikhlinin2006} (Vikh) and
\protect\cite{bulbul2010} (Poly) models. The black line is $y=x$, unrelaxed clusters are shown in red,
 and the blue data points correspond to relaxed clusters.}
\label{fig:poly_vs_vikh}
\end{figure*}

\begin{table}
\centering
\caption{Systematic Uncertainties and Effects on $f_{\textrm{gas}}$}
\begin{tabular}{lcc}
\hline
\hline
\multirow{2}{*}{Source} & \multicolumn{2}{c}{Effect on $f_{\textrm{gas}}$}\\
       & $f_{\textrm{gas}}(r_{2500})$ & $f_{\textrm{gas}}(r_{500})$\\
\hline
\chandra\ Instrument Calibration &\\
--Surface Brightness & \multicolumn{2}{c}{$\pm{1\%}$}\\
--Temperature & \multicolumn{2}{c}{$\pm{10\%}$}\\
\hline
Hydrostatic Equilibrium & $-8\%$ & $-10\%$\\
Spherical Symmetry Assumption & $\pm{6\%}$ & $\pm{8\%}$\\
Modelling of X-ray Data & $\pm{6\%}$ & $\pm{1\%}$\\
Clumping of Gas &$\cdots$ & $-10\%$\\
\hline
\end{tabular}
\label{tab:uncertainty}
\end{table}

\section{Comparison with Previous Studies}
\label{sec:comparison}
\subsection{Average gas mass fractions from Vikhlinin et al. 2006 and Arnaud et al. 2007}
In this section we compare our result to previous studies of $f_{\textrm{gas}}$ using
hydrostatic mass estimates. We used the results from the samples
of relaxed clusters by \citet[hereafter V06]{vikhlinin2006}
and \citet[hereafter A07]{arnaud2007}, and compare those results with 
our 13 relaxed clusters. 
V06 measured $f_{\textrm{gas}}$ out to $r_{500}$ for 10 relaxed clusters 
observed with \chandra\ spanning a redshift range
$z=0.02-0.23$, with a weighted average for these 10 clusters of
\begin{equation*}
f_{\textrm{gas},\,\textrm{V}06}=0.105\pm{0.002}.
\end{equation*} 
The A07 sample consists of 10 relaxed clusters with $z \leq 0.15$
observed by \emph{XMM-Newton},
with a weighted average for the 10 clusters of
\begin{equation*}
f_{\textrm{gas},\,\textrm{A}07}=0.106\pm{0.004}.
\end{equation*}
The sample used in this work are the most luminous
clusters from the BCS and eBCS in the redshift range
$0.15~\!\!\leq\!\!~z~\!\!\leq\!\!~0.30$ \citep{dahle2006},
with a weighted average for the 13 relaxed clusters of
\begin{equation*}
f_{\textrm{gas},\,\textrm{relaxed}} = 0.150\pm{0.004}.
\end{equation*} 
Notice that the value reported in Table~\ref{tab:results} ($f_{\textrm{gas},\,\textrm{relaxed}}=0.155 \pm 0.026$)
is the value of the gas mass fraction for the average cluster profile, i.e., obtained from the
combination of all Markov chains. In this section we use the weighted average of the 13
measurements for the relaxed clusters, since this number can be compared directly with the averages
obtained from the data published in the V06 and A07 papers.

As an initial comparison, we checked for clusters used in our sample that were also analysed 
by V06 and A07. Clusters that overlap with our sample were \textit{Abell~2390}
(analysed in V06) and \textit{Abell~2204} (analysed in A07). 
A07 found $f_{\textrm{gas}}(r_{500})=0.126\pm{0.013}$ for \textit{Abell~2204},
and we measure a value of $0.163\pm{0.010}$.
Our temperature profiles do not agree,
and comparison between the A07 result shown in \cite{pointecouteau2005} and our Figure \ref{fig:T-fits-few}
indicate that our temperature profile has lower values at large radii.
Since the temperature of the cluster is the dominant factor in calculating hydrostatic
masses, the discrepancy in temperature profiles can be responsible for  the different gas mass fractions. 
For \textit{Abell~2390}, V06 calculates $f_{\textrm{gas}}(r_{500})=0.141\pm{0.009}$.
Our temperature profile agrees well with that of V06, and  our measurement 
$f_{\textrm{gas}}(r_{500})=0.131\pm{0.024}$ is also in very good agreement with V06. 

\subsection{Comparison with the Vikhlinin et al. 2006 results in a matching mass range}
\begin{figure}
\centering
\includegraphics[angle=-90,width=2.9in]{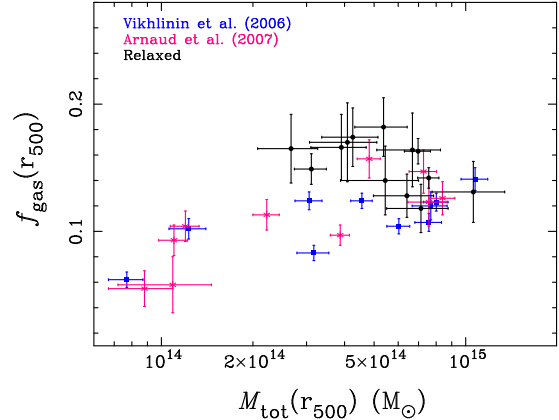}
\caption{$f_{\textrm{gas}}-M$ plot for \protect\cite{vikhlinin2006}, \protect\cite{arnaud2007},
and the relaxed clusters from this work. The mass ranges show overlap between the samples.}
\label{fig:cdf}
\end{figure}

We also compared the masses of our cluster sample with those in the V06 and A07 samples. 
As seen in Figure \ref{fig:cdf}, our clusters are generally more massive
than those in the V06 or A07 samples, yet there is an overlap in the range of masses
for all three samples. To compare clusters in a similar mass
range, we grouped clusters in two bins: bin~1 includes
clusters in the mass range 
 $2 < M_{\textrm{tot}} < 5 \times 10^{14} \textrm{M}_{\odot}$, and bin~2 with 
 $M_{\textrm{tot}} > 5 \times 10^{14} \textrm{M}_{\odot}$.
The weighted average of $f_{\textrm{gas}}$  for these bins are reported
in Table~\ref{tab:mass_binned}, showing that 
there is a significant difference between our $f_{\textrm{gas}}$ measurements
and those of V06 and A07,
especially for  bin~1. Also, contrary to V06 and A07, 
our results do not show an increase in $f_{\textrm{gas}}$ with mass between the two mass bins.

To further investigate this disagreement, we decided to analyse all of the V06 clusters
that did not require \textit{ROSAT} data for the
measurement of the background, namely A133, A1413, A383, and A907 \citep{vikhlinin2006}. 
We extracted temperature profiles and
obtained surface brightness profiles for these four clusters, and then measured masses,
using the same reduction and analysis procedure as for all other clusters in this work. 
Our weighted average for the four clusters is $0.114\pm{0.006}$ (see Table \ref{tab:V06comp}). 
This is in agreement with the results from V06,
who measure a value of $0.109\pm{0.003}$ for these clusters. We therefore conclude that, on average,
our method of analysis yields statistically consistent results to those of V06.

Of the four clusters in this comparison sample, three are in agreement at the 1$\sigma$ level.
The only cluster we do not find statistical agreement with is A1413:
we measure $f_{\textrm{gas}}(r_{500})=0.161\pm{0.011}$, whereas V06 reports $0.107\pm{0.007}$.
Since the temperature we obtain for A1413 at large radii is lower than that of V06, 
 we conclude that this is likely the reason for
the disagreement, since a lower temperature would result in a lower total mass and thus a higher gas mass fraction.
 Differences in the \chandra\ calibration or other
aspects of the data analysis are likely responsible for the disagreement between our results and
those of V06 for A1413.
We also note that A07 also analysed A1413 and measured $f_{\textrm{gas}}(r_{500})=0.157\pm{0.015}$, 
which is in very good agreement with our value.

\subsection{A possible luminosity--selection bias for the gas mass fraction}
The sample of 35 clusters used in this work was selected as the most 
X-ray luminous in the $0.15-0.30$ redshift range.
We calculated the X-ray luminosity in the $0.6-9.0$ keV band
using spectra within the $(0.15-1)\,r_{500}$ region \citep{giles2012}.
We compare the $L_{\textrm{X}}$ for the five relaxed clusters in bin~1 of our sample, and
the three clusters (A133, A383, and A907) from the V06 sample in the same mass range.
 A133, A383, and A907 were found to have luminosity of respectively
 $5.5\times10^{43}$, 
$2.8\times10^{44}$, and $3.8\times10^{44}$ erg s$^{-1}$, for an average of $2.4 \times10^{44}$~erg s$^{-1}$,
a factor of approximately \emph{three times lower} than the values for the five clusters in the same
mass range present in our sample (Table~\ref{tab:mass_binned}).
This analysis of clusters in the same mass range indicates that 
the selection of clusters based on X-ray luminosity -- as in the
case of our sample -- may result in the preferential selection of the high-$f_{\textrm{gas}}$
tail of the cluster $f_{\textrm{gas}}$ distribution for a given mass. This is not surprising,
since both $f_{\textrm{gas}}$ and $L_{\textrm{X}}$ depend on the gas mass content of the cluster.
This conclusion is also supported by the scaling relations measured by \cite{giles2012}, in which we 
use the same sample of high-$L_{\textrm{X}}$ clusters to find that there is an offset with
respect to earlier studies that can be explained by a higher gas mass for a fixed total mass.

\begin{table*}
\centering
\caption{Cluster Properties for Mass Bin~1 ( $2 < M_{\textrm{tot}} < 5 \times 10^{14} \,\textrm{M}_{\odot}$) 
and Mass Bin~2 ($M_{\textrm{tot}} > 5 \times 10^{14} \,\textrm{M}_{\odot}$)}
\label{tab:mass_binned}
\begin{tabular}{lccccc}
\hline
\hline \\[-0.75em]
\multirow{2}{*}{Sample} & \multicolumn{2}{c}{$f_{\textrm{gas}}(r_{500})$} & \multicolumn{2}{c}{Number of Clusters} & $L_{\textrm{X}}$\\[0.15em]
       & Bin 1 & Bin 2 & \hspace{1em}Bin 1 & Bin 2 & Bin 1 ($10^{44}$ erg s$^{-1}$)\\[0.25em]
\hline \\[-0.75em]
\protect\cite{vikhlinin2006} & $0.109\pm{0.004}$ & $0.116\pm{0.003}$ & \hspace{1em}3 & 5 & 2.4\\
\protect\cite{arnaud2007} & $0.111\pm{0.006}$ & $0.131\pm{0.009}$ & \hspace{1em}3 & 3 & \nodata\\
Relaxed Clusters & $0.158\pm{0.009}$ & $0.147\pm{0.005}$ & \hspace{1em}5 & 8 & 7.1\\
\hline \\[-0.75em]
\end{tabular}
\end{table*}

\begin{table*}
\centering
\caption{Comparison of Clusters with Vikhlinin et al. (2006)}
\label{tab:V06comp}
\begin{tabular}{lcccccc}
\hline
\hline \\[-0.75em]
\multirow{2}{*}{Cluster} & $r_{500}$ & $M_{\textrm{tot}}(r_{2500})$ & \multirow{2}{*}{$f_{\textrm{gas}}(r_{2500})$} & $M_{\textrm{tot}}(r_{500})$ & \multirow{2}{*}{$f_{\textrm{gas}}(r_{500})$} & Weighted Avg.\\
         & (kpc) & ($10^{14} \,\textrm{M}_{\odot}$) & & ($10^{14} \,\textrm{M}_{\odot}$) & & $f_{\textrm{gas}}(r_{500})$\\[0.25em]
\hline \\[-0.75em]
\multicolumn{7}{c}{Vikhlinin et al. (2006)}\\[0.5em]
A133 & $1007\pm{41}$ & $1.13\pm{0.07}$ & $0.067\pm{0.002}$ & $3.17\pm{0.38}$ & $0.083\pm{0.006}$ & \multirow{4}{*}{$0.109\pm{0.003}$}\\
A1413 & $1299\pm{43}$ & $3.01\pm{0.18}$ & $0.094\pm{0.003}$ & $7.57\pm{0.76}$ & $0.107\pm{0.007}$ &\\
A383 & $944\pm{32}$ & $1.64\pm{0.14}$ & $0.092\pm{0.005}$ & $3.06\pm{0.31}$ & $0.124\pm{0.007}$ & \\
A907 & $1096\pm{30}$ & $2.21\pm{0.14}$ & $0.091\pm{0.003}$ & $4.56\pm{0.37}$ & $0.124\pm{0.006}$ &\\
\hline \hline \\[-0.75em]
\multicolumn{7}{c}{Landry et al. (2012)}\\[0.5em]
A133 & $1027\pm{48}$ & $1.11\pm{0.07}$ & $0.066\pm{0.002}$ & $3.26\pm{0.46}$ & $0.080\pm{0.008}$ & \multirow{4}{*}{$0.114\pm{0.006}$}\\
A1413 & $1160\pm{40}$ & $3.55\pm{0.45}$ & $0.097\pm{0.007}$ & $5.06\pm{0.52}$ & $0.161\pm{0.011}$ &\\
A383 & $886\pm{98}$ & $1.40\pm{0.46}$ & $0.120\pm{0.022}$ & $2.36\pm{0.79}$ & $0.169\pm{0.045}$ &\\
A907 & $1142\pm{53}$ & $2.19\pm{0.12}$ & $0.104\pm{0.003}$ & $4.91\pm{0.67}$ & $0.132\pm{0.013}$ &\\
\hline \\[-0.75em]
\end{tabular}
\end{table*}

\section{Discussion and Conclusions}
\label{sec:conclusion}
In this paper we studied the gas mass fraction for the 35 most
luminous clusters in the BCS/eBCS  at redshift $z=0.15-0.30$ \citep{dahle2006}.
In accord with earlier studies, we find that the gas mass fraction increases with radius, 
and thus the value of $f_{\textrm{gas}}$ depends
on the radius used. 

Low gas mass fractions (e.g., $f_{\textrm{gas}}\leq 0.1$) 
have been observed in groups of galaxies \citep{sun2009},
likely because the gas is not fully bound to the group 
due to the shallow gravitational potential.
Recent studies 
report that the cluster baryon fraction falls short of the cosmic baryon 
fraction at $r_{500}$ even in more massive clusters 
\citep[e.g.,][]{vikhlinin2006,arnaud2007,ettori2009,rasheed2010}.
In our sample of the most X-ray luminous clusters
at redshift $z=0.15-0.30$, we find a significantly higher 
 average value of the gas mass fraction at $r_{500}$, $f_{\textrm{gas}} = 0.163\pm{0.032}$. 
To  find the total baryon fraction we used the stellar fraction
 from \cite{giodini2009} and \cite{gonzalez2007}.
Comparing the total baryon fraction with the cosmic baryon fraction 
($\Omega_{\textrm{b}} / \Omega_{\textrm{M}} = 0.167\pm{0.006}$)
we find that the two measurements agree at the 1$\sigma$ level, i.e., the gas mass fraction at $r_{500}$
is in fact consistent with the cosmological value of $\Omega_{\textrm{b}} / \Omega_{\textrm{M}}$, and there are no
\emph{missing baryons} within $r_{500}$ in the most luminous and massive clusters.
A recent study by \cite{miller2012} also find a gas mass fraction in agreement with the cosmic
baryon fraction using data from \chandra, \textit{XMM-Newton}, and \textit{Suzaku}.

One question that remains open is: what happens to $f_{\textrm{gas}}$ beyond $r_{500}$? As seen in
Figure \ref{fig:vikh-fgas-profile-sample}, the gas mass fraction increases with radius.
A naive extrapolation beyond $r_{500}$ will give an even higher value for the gas mass fraction. The total
baryon fraction in this case will be greater than the cosmological value, $\Omega_{\textrm{b}} / \Omega_{\textrm{M}}$. High values of the gas mass fractions towards the virial radius have been
 reported in recent studies based on \emph{Suzaku} data \citep[e.g.,][]{simionescu2011}.
However, there are two important reasons why simple extrapolations to large radii
of the measured radial trend of $f_{\textrm{gas}}(r)$ may not be valid. First, the gas
beyond $r_{500}$ is not likely to be in hydrostatic equilibrium. Non-thermal pressure support
will become more significant at radii $>r_{500}$ and will impact the measurement of the total
mass \citep{nagai2007b,lau2009}. 
An increase of non-thermal pressure support would cause
an underestimate of total mass, and thus an overestimate of $f_{\textrm{gas}}$. 
Therefore, the true gas mass fraction
would be lower than our measured value in the presence of non-thermal pressure.
The second reason is that clumping of the gas may also become more
important at large radii.
Observations and simulations show that the gas clumping factor increases beyond $r_{500}$ and will
considerably bias X-ray mass measurements \citep{nagai2011,simionescu2011,eckert2012}.
As a result of clumping, the true gas mass fraction at large radii would be lower than the 
value measured assuming a smooth distribution of the gas.

We also consider the effect that our sample selection criteria
could have on the measurement of $f_{\textrm{gas}}$. The  sample 
presented in this paper comprises clusters  selected
because of their high X-ray flux and luminosity, i.e., our sample
preferentially selects clusters with a high $L_{\textrm{X}}$ for a given mass. 
Since $L_{\textrm{X}}$ depends on  the gas mass, a luminosity-selected 
sample may have  a larger fraction of clusters with high $f_{\textrm{gas}}$
than samples selected with other criteria.
The extent of any such bias would depend on the correlation between the
scatter in $f_{\textrm{gas}}$ and $L_{\textrm{X}}$ at fixed mass, and the magnitude of the
intrinsic scatter in $f_{\textrm{gas}}$. Full account of this effect requires
modelling of the selection function and population scatter of the
sample, which is presented in \cite{giles2012}. 
For the current
study, we estimate the intrinsic scatter of $f_{\textrm{gas}}$ in our sample as
$11\%\pm{4}\%$. This may be underestimated if clusters from the low $f_{\textrm{gas}}$ tail
of the distribution are excluded by the luminosity selection of the
sample.
If this bias exists, then it could pose problems for using the
gas mass fraction for cosmology. In fact, clusters used for cosmological applications
are generally selected on the basis of their high luminosity, allowing
their study at high redshift. Such clusters would be biased towards high
$f_{\textrm{gas}}$, potentially distorting the cosmological constraints derived
from the $f_{\textrm{gas}}(z)$ tests. The implication is that complete (or at
least statistically representative) samples should be used for the
$f_{\textrm{gas}}(z)$ tests, so that the selection bias can be
corrected. While these samples would necessarily contain
morphologically disturbed clusters, our results show that $f_{\textrm{gas}}$
measurements are not significantly affected by cluster morphology,
at least out to $r_{500}$;
\cite{giles2012} also find an agreement between scaling
relations for relaxed and unrelaxed hydrostatic masses. This
suggests that complete samples of clusters could be used for
$f_{\textrm{gas}}(z)$ tests.
A complementary approach would be to study the baryon fraction of
clusters selected independently of the ICM (e.g., red sequence or
weak lensing selected clusters). This would give a measurement of
the range of baryon fractions that is free from selection biases
associated with their X-ray emission.

We conclude that the large value of $f_{\textrm{gas}}$ at $r_{500}$ measured
for this sample is representative of the high-$f_{\textrm{gas}}$ tail
of the cluster population, and that the most massive and X-ray luminous clusters
in this redshift range have the cosmological ratio of baryons to dark matter even
at $r_{500}$.

\clearpage
\newpage

\begin{table*}\scriptsize
\caption{\protect\cite{vikhlinin2006} Model Density Parameters}
\begin{tabular}{lccccccccc}
\hline
\hline
Cluster & $n_{\textrm{e0}}$ & $r_{\textrm{c}}$ & $\beta$ & $r_{\textrm{s}}$ & $\varepsilon$ & $n_{\textrm{e02}}$ & $\gamma$ & $\alpha$\\
        & (10$^{-2}$ cm$^{-3}$) & (arcsec) & & (arcsec) & & & & &\\
\hline
A115 & 5.86 $\pm^{0.72}_{0.77}$ & 4.89 $\pm^{0.72}_{0.60}$ & 0.42 $\pm^{0.01}_{0.01}$ & 751.40 $\pm^{305.30}_{242.50}$ & 2.01 $\pm^{1.15}_{1.14}$ & 0.0 & 3.0 & 0.0\\
A1423 & 2.58 $\pm^{0.19}_{0.17}$ & 9.85 $\pm^{0.82}_{0.83}$ & 0.44 $\pm^{0.01}_{0.01}$ & 280.80 $\pm^{110.30}_{110.90}$ & 0.90 $\pm^{0.63}_{0.40}$ & 0.0 & 3.0 & 0.0\\
A1576 & 1.16 $\pm^{0.14}_{0.12}$ & 23.33 $\pm^{3.21}_{3.13}$ & 0.53 $\pm^{0.02}_{0.02}$ & 100.0 & 0.0 & 0.0 & 3.0 & 0.0\\
A1682 & 0.64 $\pm^{0.10}_{0.07}$ & 20.19 $\pm^{6.17}_{6.06}$ & 0.33 $\pm^{0.04}_{0.05}$ & 190.80 $\pm^{47.41}_{55.52}$ & 2.33 $\pm^{0.67}_{0.54}$ & 0.0 & 3.0 & 0.0\\
A1758 & 0.37 $\pm^{0.01}_{0.01}$ & 81.45 $\pm^{2.14}_{2.14}$ & 0.7 & 100.0 & 0.0 & 0.0 & 3.0 & 0.0\\
A1763 & 0.81 $\pm^{0.03}_{0.03}$ & 35.77 $\pm^{2.74}_{2.63}$ & 0.48 $\pm^{0.02}_{0.02}$ & 425.10 $\pm^{78.03}_{49.17}$ & 3.0 & 0.0 & 3.0 & 0.0\\
A1835 & 9.84 $\pm^{0.17}_{0.17}$ & 6.69 $\pm^{0.14}_{0.13}$ & 0.50 $\pm^{0.00}_{0.00}$ & 123.10 $\pm^{8.79}_{7.47}$ & 1.18 $\pm^{0.07}_{0.07}$ & 0.0 & 3.0 & 0.0\\
A1914 & 1.47 $\pm^{0.05}_{0.05}$ & 54.80 $\pm^{2.55}_{2.43}$ & 0.73 $\pm^{0.02}_{0.02}$ & 500.0 & 0.0 & 0.0 & 3.0 & 0.0\\
A2111 & 0.60 $\pm^{0.02}_{0.02}$ & 42.10 $\pm^{2.31}_{2.43}$ & 0.54 $\pm^{0.01}_{0.02}$ & 10.0 & 0.0 & 0.0 & 3.0 & 0.0\\
A2204 & 30.10 $\pm^{1.54}_{1.73}$ & 3.71 $\pm^{0.19}_{0.16}$ & 0.46 $\pm^{0.00}_{0.00}$ & 133.00 $\pm^{9.62}_{8.80}$ & 1.12 $\pm^{0.06}_{0.06}$ & 0.0 & 3.0 & 0.0\\
A2219 & 0.90 $\pm^{0.04}_{0.04}$ & 59.85 $\pm^{3.51}_{3.53}$ & 0.63 $\pm^{0.02}_{0.02}$ & 500.0 & 0.0 & 0.0 & 3.0 & 0.0\\
A2261 & 3.10 $\pm^{0.15}_{0.17}$ & 13.44 $\pm^{2.53}_{1.43}$ & 0.44 $\pm^{0.06}_{0.04}$ & 70.00 $\pm^{44.92}_{14.46}$ & 0.98 $\pm^{0.15}_{0.17}$ & 0.0 & 3.0 & 0.0\\
A2390 & 7.64 $\pm^{1.45}_{1.06}$ & 3.86 $\pm^{0.77}_{0.81}$ & 0.36 $\pm^{0.01}_{0.01}$ & 145.10 $\pm^{13.73}_{12.32}$ & 2.67 $\pm^{0.27}_{0.23}$ & 0.0 & 3.0 & 0.0\\
A2552 & 1.74 $\pm^{0.20}_{0.19}$ & 9.42 $\pm^{3.38}_{2.53}$ & 0.29 $\pm^{0.06}_{0.05}$ & 51.24 $\pm^{9.26}_{8.05}$ & 1.57 $\pm^{0.26}_{0.37}$ & 0.0 & 3.0 & 0.0\\
A2631 & 0.57 $\pm^{0.02}_{0.02}$ & 51.52 $\pm^{8.06}_{6.09}$ & 0.55 $\pm^{0.09}_{0.07}$ & 178.90 $\pm^{77.79}_{40.74}$ & 2.0 & 0.0 & 3.0 & 0.0\\
A267 & 0.93 $\pm^{0.05}_{0.05}$ & 35.98 $\pm^{2.70}_{2.61}$ & 0.64 $\pm^{0.02}_{0.03}$ & 100.0 & 0.0 & 0.0 & 3.0 & 0.0\\
A520 & 0.37 $\pm^{0.01}_{0.01}$ & 127.30 $\pm^{7.30}_{6.53}$ & 0.85 $\pm^{0.04}_{0.04}$ & 900.0 & 0.0 & 0.0 & 3.0 & 0.0\\
A586 & 1.63 $\pm^{0.04}_{0.04}$ & 31.30 $\pm^{0.96}_{0.98}$ & 0.61 $\pm^{0.01}_{0.01}$ & 10.0 & 0.0 & 0.0 & 3.0 & 0.0\\
A611 & 2.14 $\pm^{0.11}_{0.10}$ & 16.59 $\pm^{1.16}_{1.10}$ & 0.56 $\pm^{0.02}_{0.02}$ & 426.30 $\pm^{134.30}_{107.50}$ & 5.0 & 0.0 & 3.0 & 0.0\\
A665 & 1.00 $\pm^{0.05}_{0.05}$ & 23.04 $\pm^{1.63}_{1.52}$ & 0.41 $\pm^{0.01}_{0.01}$ & 472.20 $\pm^{23.93}_{22.77}$ & 5.0 & 0.0 & 3.0 & 0.0\\
A68 & 0.82 $\pm^{0.13}_{0.08}$ & 43.14 $\pm^{9.30}_{15.41}$ & 0.62 $\pm^{0.10}_{0.19}$ & 233.40 $\pm^{110.30}_{104.50}$ & 2.98 $\pm^{1.58}_{1.76}$ & 0.0 & 3.0 & 0.0\\
A697 & 0.90 $\pm^{0.07}_{0.07}$ & 43.42 $\pm^{6.02}_{6.59}$ & 0.58 $\pm^{0.04}_{0.07}$ & 279.30 $\pm^{159.40}_{162.50}$ & 1.04 $\pm^{0.83}_{0.72}$ & 0.0 & 3.0 & 0.0\\
A773 & 0.87 $\pm^{0.02}_{0.02}$ & 45.61 $\pm^{1.23}_{1.21}$ & 0.62 $\pm^{0.01}_{0.01}$ & 100.0 & 0.0 & 0.0 & 3.0 & 0.0\\
A781 & 0.27 $\pm^{0.02}_{0.02}$ & 111.40 $\pm^{4.74}_{4.52}$ & 0.9 & 500.0 & 0.0 & 0.0 & 3.0 & 0.0\\
A963 & 2.57 $\pm^{0.14}_{0.14}$ & 12.66 $\pm^{1.17}_{1.09}$ & 0.43 $\pm^{0.01}_{0.01}$ & 176.10 $\pm^{18.50}_{20.74}$ & 3.05 $\pm^{0.33}_{0.43}$ & 0.0 & 3.0 & 0.0\\
MS 1455.0+2232 & 7.30 $\pm^{0.37}_{0.34}$ & 11.47 $\pm^{0.68}_{0.71}$ & 0.45 $\pm^{0.00}_{0.00}$ & 4.96 $\pm^{0.85}_{0.62}$ & 1.0 & 0.0 & 3.0 & 0.0\\
RX J0437.1+0043 & 4.07 $\pm^{0.37}_{0.29}$ & 6.48 $\pm^{0.91}_{0.93}$ & 0.38 $\pm^{0.03}_{0.03}$ & 48.88 $\pm^{9.53}_{7.39}$ & 1.35 $\pm^{0.16}_{0.15}$ & 0.0 & 3.0 & 0.0\\
RX J0439.0+0715 & 3.06 $\pm^{0.73}_{0.64}$ & 8.69 $\pm^{4.49}_{2.90}$ & 0.38 $\pm^{0.06}_{0.06}$ & 108.40 $\pm^{47.30}_{31.81}$ & 2.29 $\pm^{0.51}_{0.41}$ & 0.0 & 3.0 & 0.0\\
RX J1720.1+2638 & 7.37 $\pm^{0.26}_{0.22}$ & 9.38 $\pm^{0.54}_{0.50}$ & 0.48 $\pm^{0.01}_{0.01}$ & 97.82 $\pm^{19.53}_{13.87}$ & 0.92 $\pm^{0.07}_{0.07}$ & 0.0 & 3.0 & 0.0\\
RX J2129.6+0005 & 6.43 $\pm^{0.43}_{0.38}$ & 6.72 $\pm^{0.67}_{0.62}$ & 0.44 $\pm^{0.02}_{0.02}$ & 76.71 $\pm^{21.93}_{15.13}$ & 1.12 $\pm^{0.15}_{0.12}$ & 0.0 & 3.0 & 0.0\\
Z2089 & 11.21 $\pm^{1.92}_{1.55}$ & 5.46 $\pm^{1.17}_{1.18}$ & 0.54 $\pm^{0.03}_{0.05}$ & 99.43 $\pm^{182.70}_{51.51}$ & 1.0 & 0.0 & 3.0 & 0.0\\
Z3146 & 10.49 $\pm^{0.33}_{0.34}$ & 6.89 $\pm^{0.27}_{0.27}$ & 0.50 $\pm^{0.01}_{0.01}$ & 96.05 $\pm^{10.79}_{9.96}$ & 1.56 $\pm^{0.18}_{0.17}$ & 0.0 & 3.0 & 0.0\\
Z5247 & 0.19 $\pm^{0.01}_{0.01}$ & 84.60 $\pm^{6.46}_{5.82}$ & 0.5 & 100.0 & 0.0 & 0.0 & 3.0 & 0.0\\
Z5768 & 0.63 $\pm^{0.15}_{0.14}$ & 15.49 $\pm^{5.89}_{3.74}$ & 0.45 $\pm^{0.05}_{0.04}$ & 100.0 & 0.0 & 0.0 & 3.0 & 0.0\\
Z7215 & 0.62 $\pm^{0.07}_{0.06}$ & 51.41 $\pm^{9.43}_{7.84}$ & 0.76 $\pm^{0.09}_{0.07}$ & 500.0 & 0.0 & 0.0 & 3.0 & 0.0\\

\hline
\end{tabular}
\label{tab:vikh_density}
\end{table*}

\clearpage
\newpage

\begin{table*}\scriptsize
\centering
\caption{\protect\cite{vikhlinin2006} Model Temperature Parameters}
\begin{tabular}{lccccccccc}
\hline
\hline
Cluster & $T_{0}$ & $T_{\textrm{min}}$ & $r_{\textrm{cool}}$ & $a_{\textrm{cool}}$ & $r_{\textrm{t}}$ & $a$ & $b$ & $c$ & $\chi^{2}_{\textrm{tot}} \textrm{(d.o.f.)}$\\
        & (keV) & (keV) & (arcsec) & & (arcsec) & & & &\\
\hline
A115 & 21.94 $\pm^{7.07}_{5.74}$ & 2.08 $\pm^{0.36}_{0.39}$ & 222.50 $\pm^{71.81}_{78.68}$ & 1.0 & 307.60 $\pm^{82.31}_{72.12}$ & 0.0 & 2.0 & 2.0 & 92.50 (128)\\
A1423 & 7.47 $\pm^{1.69}_{1.26}$ & 4.98 $\pm^{1.03}_{1.06}$ & 50.0 & 2.0 & 183.80 $\pm^{97.22}_{66.96}$ & 0.0 & 2.0 & 1.0 & 56.65 (113)\\
A1576 & 9.20 $\pm^{2.48}_{1.69}$ & 9.20 & 50.0 & 1.0 & 411.50 $\pm^{373.90}_{260.90}$ & 0.0 & 1.0 & 1.0 & 16.80 (74)\\
A1682 & 8.91 $\pm^{2.55}_{1.73}$ & 2.0 & 16.43 $\pm^{8.94}_{8.94}$ & 2.0 & 121.70 $\pm^{71.64}_{40.70}$ & 0.0 & 2.0 & 1.0 & 52.65 (89)\\
A1758 & 10.40 $\pm^{2.47}_{2.34}$ & 10.40 & 10.0 & 0.0 & 100.0 & 0.0 & 1.0 & 0.0 & 104.51 (59)\\
A1763 & 8.97 $\pm^{1.03}_{1.03}$ & 8.97 & 50.0 & 1.0 & 1200.0 & 0.0 & 1.0 & 1.0 & 97.75 (126)\\
A1835 & 21.98 $\pm^{9.31}_{5.26}$ & 3.0 & 51.01 $\pm^{23.25}_{13.88}$ & 1.0 & 307.10 $\pm^{256.50}_{80.36}$ & 0.0 & 1.14 $\pm^{0.52}_{0.34}$ & 2.0 & 196.64 (83)\\
A1914 & 11.38 $\pm^{1.91}_{1.27}$ & 11.38 & 50.0 & 1.0 & 487.00 $\pm^{350.70}_{227.60}$ & 0.0 & 1.0 & 1.0 & 66.00 (109)\\
A2111 & 8.43 $\pm^{1.86}_{1.40}$ & 6.0 & 50.0 & 1.0 & 469.30 $\pm^{306.20}_{184.80}$ & 0.0 & 2.0 & 2.0 & 43.95 (93)\\
A2204 & 9.94 $\pm^{0.59}_{0.49}$ & 0.09 $\pm^{0.10}_{0.06}$ & 7.78 $\pm^{0.69}_{0.62}$ & 2.0 & 636.70 $\pm^{329.60}_{187.00}$ & 0.0 & 1.94 $\pm^{0.57}_{0.48}$ & 2.0 & 153.68 (136)\\
A2219 & 14.16 $\pm^{2.41}_{1.87}$ & 14.16 & 50.0 & 1.0 & 305.60 $\pm^{230.50}_{120.80}$ & 0.0 & 1.0 & 1.0 & 21.66 (104)\\
A2261 & 11.66 $\pm^{1.72}_{1.39}$ & 4.0 & 30.0 & 0.5 & 299.80 $\pm^{117.60}_{80.10}$ & 0.0 & 2.0 & 2.0 & 73.71 (105)\\
A2390 & 30.11 $\pm^{4.55}_{4.27}$ & 0.0 & 60.0 & 0.61 $\pm^{0.12}_{0.13}$ & 1000.0 & 0.0 & 1.0 & 4.00 $\pm^{1.59}_{1.30}$ & 87.44 (152)\\
A2552 & 8.77 $\pm^{1.17}_{1.23}$ & 8.77 & 10.0 & 0.0 & 300.0 & 0.0 & 2.0 & 2.0 & 61.66 (74)\\
A2631 & 9.36 $\pm^{1.97}_{1.43}$ & 9.36 & 10.0 & 0.0 & 316.80 $\pm^{239.40}_{149.70}$ & 0.0 & 1.0 & 1.0 & 29.95 (74)\\
A267 & 5.08 $\pm^{0.91}_{0.86}$ & 5.08 & 10.0 & 0.0 & 385.30 $\pm^{385.50}_{143.80}$ & 0.0 & 2.0 & 2.0 & 122.55 (152)\\
A520 & 8.76 $\pm^{0.55}_{0.58}$ & 8.76 & 3.0 & 1.0 & 346.50 $\pm^{45.52}_{37.64}$ & 0.0 & 2.0 & 2.0 & 43.71 (128)\\
A586 & 8.83 $\pm^{0.76}_{0.74}$ & 5.0 & 60.0 & 0.2 & 330.30 $\pm^{88.25}_{64.65}$ & 0.0 & 2.0 & 2.0 & 49.60 (103)\\
A611 & 9.86 $\pm^{1.33}_{1.57}$ & 5.0 & 20.0 & 1.0 & 300.0 & 0.0 & 2.0 & 2.0 & 26.61 (47)\\
A665 & 11.71 $\pm^{2.86}_{2.02}$ & 6.41 $\pm^{1.17}_{1.27}$ & 100.0 & 1.0 & 510.40 $\pm^{206.40}_{155.30}$ & 0.0 & 2.0 & 2.0 & 180.08 (179)\\
A68 & 10.08 $\pm^{3.95}_{2.06}$ & 10.08 & 50.0 & 1.0 & 269.60 $\pm^{450.00}_{161.60}$ & 0.0 & 1.0 & 1.0 & 23.82 (78)\\
A697 & 13.87 $\pm^{3.86}_{2.52}$ & 6.0 & 38.38 $\pm^{29.42}_{22.40}$ & 2.0 & 407.10 $\pm^{334.20}_{208.70}$ & 0.0 & 1.0 & 1.0 & 29.05 (117)\\
A773 & 7.73 $\pm^{0.81}_{0.66}$ & 7.73 & 10.0 & 0.0 & 1000.0 & 0.0 & 1.0 & 1.0 & 127.68 (94)\\
A781 & 6.53 $\pm^{1.13}_{1.12}$ & 6.53 & 50.0 & 0.0 & 1000.0 & 0.0 & 1.0 & 1.0 & 34.04 (75)\\
A963 & 10.65 $\pm^{2.04}_{1.35}$ & 3.0 & 20.0 & 1.0 & 98.05 $\pm^{26.23}_{30.87}$ & 0.0 & 2.0 & 1.0 &76.59 (77)\\
MS 1455.0+2232 & 4.91 $\pm^{0.38}_{0.42}$ & 3.87 $\pm^{0.66}_{0.63}$ & 20.0 & 2.0 & 2000.0 & 0.0 & 2.0 & 2.0 &150.82 (102)\\
RX J0437.1+0043 & 10.38 $\pm^{2.08}_{1.63}$ & 3.28 $\pm^{1.13}_{1.38}$ & 30.0 & 1.0 & 278.90 $\pm^{197.90}_{102.60}$ & 0.0 & 2.0 & 2.0 & 49.92 (57)\\
RX J0439.0+0715 & 11.54 $\pm^{2.15}_{1.77}$ & 3.0 & 50.0 & 1.0 & 190.50 $\pm^{54.50}_{50.20}$ & 0.0 & 2.0 & 2.0 & 14.34 (90)\\
RX J1720.1+2638 & 13.94 $\pm^{2.77}_{2.14}$ & 4.37 $\pm^{0.41}_{0.44}$ & 73.39 $\pm^{13.89}_{12.59}$ & 2.0 & 363.80 $\pm^{237.70}_{145.00}$ & 0.0 & 1.0 & 1.0 & 163.51 (101)\\
RX J2129.6+0005 & 9.43 $\pm^{2.19}_{1.43}$ & 4.19 $\pm^{0.71}_{0.82}$ & 32.80 $\pm^{17.64}_{14.19}$ & 2.0 & 141.70 $\pm^{70.04}_{41.80}$ & 0.0 & 2.0 & 1.0 & 44.31 (74)\\
Z2089 & 4.86 $\pm^{0.98}_{0.85}$ & 3.5 & 50.0 & 0.70 $\pm^{0.63}_{0.49}$ & 942.70 $\pm^{698.10}_{688.10}$ & 0.0 & 2.0 & 2.0 & 6.93 (43)\\
Z3146 & 10.39 $\pm^{4.10}_{1.85}$ & 2.0 & 18.84 $\pm^{20.45}_{8.12}$ & 1.0 & 265.40 $\pm^{165.40}_{92.45}$ & 0.0 & 2.0 & 2.0 & 98.22 (90)\\
Z5247 & 7.15 $\pm^{2.41}_{1.57}$ & 7.15 & 50.0 & 0.0 & 241.50 $\pm^{328.00}_{129.00}$ & 0.0 & 1.0 & 1.0 & 43.82 (82)\\
Z5768 & 4.55 $\pm^{1.10}_{1.17}$ & 4.55 & 50.0 & 0.0 & 500.0 & 0.0 & 1.0 & 1.0 & 29.35 (59)\\
Z7215 & 8.68 $\pm^{2.12}_{1.73}$ & 2.0 & 17.11 $\pm^{12.23}_{10.23}$ & 2.0 & 300.0 & 0.0 & 2.0 & 2.0 & 18.15 (59)\\

\hline
\end{tabular}
\label{tab:vikh_temperature}
\end{table*}

\clearpage
\newpage

\begin{table*}\tiny
\centering
\caption{\protect\cite{bulbul2010} Model Parameters}
\begin{tabular}{lcccccccccc}
\hline
\hline
Cluster & $n_{\textrm{e0}}$ & $r_{\textrm{s}}$ & $n$ & $\beta+1$ & $T_{0}$ & $r_{\textrm{cool}}$ & $\xi$ & $a_{\textrm{cool}}$
        & $\chi^{2}_{\textrm{tot}} \textrm{(d.o.f.)}$\\
        & (10$^{-2}$ cm$^{-3}$) & (arcsec) & & & (keV) & (arcsec) & & &\\
\hline
A115 & 0.80 $\pm^{0.16}_{0.12}$ & 22.54 $\pm^{7.27}_{5.20}$ & 6.72 $\pm^{2.33}_{1.71}$ & 2.27 $\pm^{0.12}_{0.08}$ & 10.62 $\pm^{1.00}_{0.98}$ & 29.82 $\pm^{2.63}_{2.64}$ & 0.16 $\pm^{0.03}_{0.03}$ & 2.0 & 89.66 (130)\\
A1423 & 0.60 $\pm^{0.14}_{0.14}$ & 42.07 $\pm^{26.91}_{17.53}$ & 2.08 $\pm^{0.50}_{0.44}$ & 3.28 $\pm^{0.86}_{0.44}$ & 18.50 $\pm^{3.22}_{3.24}$ & 42.83 $\pm^{10.77}_{6.65}$ & 0.17 $\pm^{0.04}_{0.04}$ & 1.39 $\pm^{0.19}_{0.22}$ & 57.13 (113)\\
A1576 & 1.81 $\pm^{0.23}_{0.16}$ & 20.77 $\pm^{4.47}_{4.70}$ & 4.91 $\pm^{1.04}_{1.20}$ & 2.53 $\pm^{0.18}_{0.11}$ & 11.95 $\pm^{2.05}_{1.52}$ & 10.0 & 1.0 & 2.0 & 14.22 (74)\\
A1682 & 0.41 $\pm^{0.08}_{0.05}$ & 272.60 $\pm^{135.90}_{130.10}$ & 4.0 & 3.95 $\pm^{0.69}_{0.69}$ & 9.39 $\pm^{0.80}_{0.80}$ & 33.84 $\pm^{13.17}_{10.14}$ & 0.56 $\pm^{0.11}_{0.07}$ & 2.0 & 63.18 (92)\\
A1758 & 0.51 $\pm^{0.02}_{0.02}$ & 347.10 $\pm^{9.29}_{9.04}$ & 10.0 & 3.0 & 11.66 $\pm^{1.20}_{2.21}$ & 10.0 & 1.0 & 2.0 & 115.65 (59)\\
A1763 & 1.17 $\pm^{0.06}_{0.06}$ & 41.05 $\pm^{5.47}_{4.18}$ & 10.0 & 2.27 $\pm^{0.02}_{0.01}$ & 9.66 $\pm^{0.93}_{0.54}$ & 10.0 & 1.0 & 1.0 & 121.18 (127)\\
A1835 & 4.52 $\pm^{0.37}_{0.47}$ & 14.63 $\pm^{1.74}_{1.18}$ & 8.99 $\pm^{1.53}_{1.64}$ & 2.29 $\pm^{0.07}_{0.05}$ & 11.91 $\pm^{0.92}_{0.73}$ & 16.11 $\pm^{0.63}_{0.43}$ & 0.44 $\pm^{0.03}_{0.04}$ & 3.58 $\pm^{0.43}_{0.39}$ & 83.43 (84)\\
A1914 & 2.27 $\pm^{0.10}_{0.10}$ & 77.65 $\pm^{12.41}_{9.79}$ & 7.96 $\pm^{4.24}_{2.30}$ & 2.56 $\pm^{0.29}_{0.21}$ & 12.08 $\pm^{1.45}_{1.13}$ & 10.0 & 1.0 & 2.0 & 108.39 (109)\\
A2111 & 0.90 $\pm^{0.04}_{0.03}$ & 42.98 $\pm^{4.65}_{4.03}$ & 34.45 $\pm^{11.97}_{16.97}$ & 2.08 $\pm^{0.08}_{0.02}$ & 7.24 $\pm^{0.55}_{0.55}$ & 2.18 $\pm^{1.53}_{0.92}$ & 1.0 & 6.04 $\pm^{1.43}_{1.61}$ & 60.46 (91)\\
A2204 & 4.15 $\pm^{0.22}_{0.26}$ & 22.56 $\pm^{1.74}_{1.61}$ & 6.60 $\pm^{1.00}_{0.79}$ & 2.39 $\pm^{0.06}_{0.06}$ & 15.07 $\pm^{1.02}_{0.77}$ & 19.83 $\pm^{0.72}_{0.56}$ & 0.17 $\pm^{0.01}_{0.01}$ & 2.0 & 123.22 (139)\\
A2219 & 1.38 $\pm^{0.07}_{0.07}$ & 82.36 $\pm^{16.16}_{10.90}$ & 4.93 $\pm^{1.86}_{0.84}$ & 2.81 $\pm^{0.26}_{0.26}$ & 14.79 $\pm^{1.22}_{1.29}$ & 10.0 & 1.0 & 2.0 & 32.25 (104)\\
A2261 & 2.84 $\pm^{0.53}_{0.59}$ & 18.29 $\pm^{5.75}_{3.27}$ & 5.19 $\pm^{1.40}_{1.30}$ & 2.48 $\pm^{0.21}_{0.12}$ & 13.53 $\pm^{2.41}_{1.64}$ & 53.57 $\pm^{19.95}_{10.70}$ & 0.68 $\pm^{0.08}_{0.10}$ & 2.0 & 75.54 (105)\\
A2390 & 1.09 $\pm^{0.11}_{0.08}$ & 155.40 $\pm^{43.66}_{39.02}$ & 3.60 $\pm^{0.74}_{0.50}$ & 3.95 $\pm^{0.69}_{0.59}$ & 19.47 $\pm^{1.30}_{1.35}$ & 21.91 $\pm^{2.14}_{2.12}$ & 0.19 $\pm^{0.02}_{0.02}$ & 2.0 & 99.80 (153)\\
A2552 & 2.24 $\pm^{0.13}_{0.12}$ & 17.51 $\pm^{2.12}_{1.84}$ & 15.0 & 2.16 $\pm^{0.01}_{0.01}$ & 9.17 $\pm^{0.71}_{0.89}$ & 10.0 & 1.0 & 2.0 & 69.04 (76)\\
A2631 & 0.76 $\pm^{0.02}_{0.02}$ & 404.00 $\pm^{159.20}_{121.30}$ & 4.56 $\pm^{1.98}_{0.88}$ & 5.39 $\pm^{1.63}_{1.68}$ & 9.82 $\pm^{1.27}_{1.22}$ & 20.0 & 1.0 & 2.0 & 30.22 (75)\\
A267 & 1.42 $\pm^{0.09}_{0.08}$ & 57.45 $\pm^{14.89}_{12.49}$ & 3.40 $\pm^{1.12}_{0.64}$ & 3.35 $\pm^{0.55}_{0.46}$ & 7.78 $\pm^{1.01}_{1.10}$ & 20.0 & 1.0 & 2.0 & 142.48 (152)\\
A520 & 0.53 $\pm^{0.01}_{0.01}$ & 496.00 $\pm^{88.54}_{89.01}$ & 11.08 $\pm^{1.60}_{1.60}$ & 3.0 & 7.38 $\pm^{0.34}_{0.32}$ & 10.0 & 1.0 & 1.0 & 115.91 (129)\\
A586 & 0.92 $\pm^{0.16}_{0.12}$ & 68.35 $\pm^{18.96}_{15.05}$ & 2.58 $\pm^{0.43}_{0.39}$ & 3.69 $\pm^{0.29}_{0.40}$ & 18.00 $\pm^{3.99}_{2.39}$ & 87.68 $\pm^{18.20}_{11.86}$ & 0.43 $\pm^{0.06}_{0.06}$ & 2.0 & 44.07 (101)\\
A611 & 1.28 $\pm^{0.25}_{0.28}$ & 37.50 $\pm^{39.89}_{11.92}$ & 4.56 $\pm^{2.96}_{1.42}$ & 2.76 $\pm^{0.90}_{0.37}$ & 14.42 $\pm^{1.38}_{1.90}$ & 36.59 $\pm^{4.99}_{4.12}$ & 0.47 $\pm^{0.07}_{0.05}$ & 2.38 $\pm^{0.85}_{0.68}$ & 28.24 (44)\\
A665 & 0.38 $\pm^{0.06}_{0.03}$ & 127.20 $\pm^{29.24}_{35.58}$ & 3.42 $\pm^{0.68}_{0.45}$ & 3.13 $\pm^{0.29}_{0.31}$ & 17.85 $\pm^{1.31}_{1.67}$ & 60.0 & 0.23 $\pm^{0.07}_{0.04}$ & 1.0 & 238.30 (180)\\
A68 & 1.14 $\pm^{0.11}_{0.09}$ & 225.90 $\pm^{90.72}_{83.98}$ & 3.52 $\pm^{1.66}_{0.69}$ & 5.35 $\pm^{1.32}_{1.58}$ & 13.05 $\pm^{2.64}_{2.33}$ & 10.0 & 1.0 & 2.0 & 24.42 (80)\\
A697 & 0.99 $\pm^{0.19}_{0.19}$ & 68.92 $\pm^{18.88}_{12.28}$ & 10.27 $\pm^{4.59}_{4.10}$ & 2.36 $\pm^{0.28}_{0.12}$ & 13.14 $\pm^{2.52}_{1.78}$ & 100.0 & 0.79 $\pm^{0.13}_{0.15}$ & 2.0 & 31.15 (119)\\
A773 & 1.33 $\pm^{0.03}_{0.03}$ & 56.17 $\pm^{3.74}_{3.58}$ & 7.11 $\pm^{0.24}_{0.23}$ & 2.5 & 9.58 $\pm^{0.39}_{0.45}$ & 10.0 & 1.0 & 2.0 & 139.93 (94)\\
A781 & 0.36 $\pm^{0.02}_{0.02}$ & 1442.00 $\pm^{375.60}_{501.80}$ & 33.81 $\pm^{8.40}_{10.97}$ & 3.0 & 6.18 $\pm^{0.80}_{0.77}$ & 10.0 & 1.0 & 2.0 & 38.88 (74)\\
A963 & 1.54 $\pm^{0.19}_{0.14}$ & 65.20 $\pm^{13.01}_{16.06}$ & 3.78 $\pm^{1.09}_{0.58}$ & 3.22 $\pm^{0.35}_{0.42}$ & 11.54 $\pm^{1.04}_{1.27}$ & 21.27 $\pm^{3.08}_{3.15}$ & 0.49 $\pm^{0.04}_{0.04}$ & 2.0 & 82.70 (77)\\
MS 1455.0+2232 & 3.27 $\pm^{0.69}_{0.62}$ & 7.45 $\pm^{1.27}_{0.81}$ & 3.19 $\pm^{0.60}_{0.38}$ & 2.68 $\pm^{0.12}_{0.11}$ & 15.79 $\pm^{1.93}_{2.52}$ & 40.92 $\pm^{2.86}_{3.13}$ & 0.31 $\pm^{0.06}_{0.05}$ & 2.10 $\pm^{0.34}_{0.17}$ & 145.54 (100)\\
RX J0437.1+0043 & 2.61 $\pm^{0.31}_{0.30}$ & 17.57 $\pm^{2.43}_{2.08}$ & 48.98 $\pm^{14.49}_{21.81}$ & 2.05 $\pm^{0.04}_{0.01}$ & 7.78 $\pm^{0.56}_{0.42}$ & 16.76 $\pm^{2.65}_{2.61}$ & 0.57 $\pm^{0.06}_{0.06}$ & 2.0 & 57.38 (58)\\
RX J0439.0+0715 & 0.72 $\pm^{0.22}_{0.21}$ & 288.90 $\pm^{148.90}_{149.10}$ & 2.45 $\pm^{0.50}_{0.33}$ & 7.47 $\pm^{2.92}_{2.79}$ & 17.08 $\pm^{6.82}_{3.46}$ & 41.93 $\pm^{46.97}_{16.69}$ & 0.16 $\pm^{0.12}_{0.11}$ & 1.04 $\pm^{0.42}_{0.35}$ & 14.04 (89)\\
RX J1720.1+2638 & 3.78 $\pm^{0.29}_{0.23}$ & 18.30 $\pm^{1.28}_{1.66}$ & 14.53 $\pm^{2.39}_{1.66}$ & 2.17 $\pm^{0.02}_{0.03}$ & 8.81 $\pm^{0.49}_{0.38}$ & 25.12 $\pm^{1.46}_{1.28}$ & 0.42 $\pm^{0.03}_{0.03}$ & 2.07 $\pm^{0.19}_{0.18}$ & 172.51 (102)\\
RX J2129.6+0005 & 1.99 $\pm^{0.35}_{0.21}$ & 25.84 $\pm^{4.77}_{3.99}$ & 4.39 $\pm^{1.15}_{0.74}$ & 2.66 $\pm^{0.19}_{0.16}$ & 13.81 $\pm^{1.64}_{1.74}$ & 23.23 $\pm^{3.01}_{2.86}$ & 0.25 $\pm^{0.04}_{0.04}$ & 1.48 $\pm^{0.16}_{0.15}$ & 46.52 (75)\\
Z2089 & 5.13 $\pm^{1.20}_{1.09}$ & 6.20 $\pm^{3.94}_{2.13}$ & 4.53 $\pm^{1.63}_{1.16}$ & 2.49 $\pm^{0.22}_{0.15}$ & 12.05 $\pm^{2.75}_{2.65}$ & 24.99 $\pm^{33.15}_{10.42}$ & 0.2 & 1.06 $\pm^{0.40}_{0.32}$ & 6.72 (43)\\
Z3146 & 5.22 $\pm^{0.32}_{0.25}$ & 19.25 $\pm^{1.15}_{1.22}$ & 13.31 $\pm^{2.93}_{3.11}$ & 2.23 $\pm^{0.07}_{0.04}$ & 8.88 $\pm^{0.46}_{0.43}$ & 15.88 $\pm^{0.56}_{0.54}$ & 0.47 $\pm^{0.03}_{0.03}$ & 3.30 $\pm^{0.34}_{0.27}$ & 90.40 (90)\\
Z5247 & 0.24 $\pm^{0.01}_{0.01}$ & 536.40 $\pm^{26.68}_{25.53}$ & 10.0 & 3.0 & 5.17 $\pm^{0.43}_{0.39}$ & 10.0 & 1.0 & 2.0 & 35.03 (82)\\
Z5768 & 0.49 $\pm^{0.09}_{0.08}$ & 82.87 $\pm^{13.23}_{11.26}$ & 5.0 & 3.0 & 4.82 $\pm^{1.04}_{0.86}$ & 10.0 & 1.0 & 2.0 & 26.63 (60)\\
Z7215 & 0.59 $\pm^{0.14}_{0.15}$ & 319.80 $\pm^{121.60}_{119.70}$ & 9.07 $\pm^{4.53}_{3.42}$ & 3.45 $\pm^{0.96}_{0.65}$ & 9.25 $\pm^{2.33}_{1.70}$ & 64.41 $\pm^{35.54}_{40.99}$ & 0.64 $\pm^{0.18}_{0.14}$ & 2.0 & 17.64 (57)\\

\hline
\end{tabular}
\label{tab:poly_parameters}
\end{table*}

\clearpage
\newpage

\begin{table*}\scriptsize
\centering
\caption{Cluster Masses}
\begin{tabular}{lcccccccccc}
\hline
\hline
& & \multicolumn{4}{c}{$\Delta = 2500$} & \multicolumn{4}{c}{$\Delta = 500$}\\
Cluster & Model & $r_{\Delta}$ & $M_{\textrm{gas}}$ & $M_{\textrm{tot}}$ & $f_{\textrm{gas}}$ & $r_{\Delta}$ & $M_{\textrm{gas}}$ & $M_{\textrm{tot}}$ & $f_{\textrm{gas}}$\\
& & (arcsec) & (10$^{13} \,\textrm{M}_{\odot}$) & (10$^{14} \,\textrm{M}_{\odot}$) & & (arcsec) & (10$^{13} \,\textrm{M}_{\odot}$) & (10$^{14} \,\textrm{M}_{\odot}$) &\\
\hline
\underline{A115}\\
& Vikh & 134.5 $\pm^{13.7}_{12.7}$ & 1.57 $\pm^{0.28}_{0.25}$ & 1.45 $\pm^{0.49}_{0.37}$ & 0.108 $\pm^{0.015}_{0.013}$ &360.4 $\pm^{21.4}_{19.2}$ & 8.65 $\pm^{0.84}_{0.71}$ & 5.58 $\pm^{1.06}_{0.84}$ & 0.155 $\pm^{0.013}_{0.012}$\\ 
& Poly & 139.7 $\pm^{4.0}_{4.2}$ & 1.68 $\pm^{0.09}_{0.09}$ & 1.62 $\pm^{0.15}_{0.14}$ & 0.103 $\pm^{0.005}_{0.004}$ &318.7 $\pm^{11.3}_{11.7}$ & 7.08 $\pm^{0.42}_{0.41}$ & 3.86 $\pm^{0.42}_{0.41}$ & 0.184 $\pm^{0.010}_{0.009}$\\
\underline{A1423}\\
& Vikh & 126.7 $\pm^{7.0}_{7.8}$ & 1.59 $\pm^{0.15}_{0.16}$ & 1.48 $\pm^{0.26}_{0.25}$ & 0.108 $\pm^{0.009}_{0.007}$ &274.0 $\pm^{18.8}_{16.8}$ & 5.73 $\pm^{0.58}_{0.49}$ & 2.99 $\pm^{0.66}_{0.52}$ & 0.192 $\pm^{0.020}_{0.019}$\\ 
& Poly & 130.8 $\pm^{4.0}_{4.3}$ & 1.69 $\pm^{0.09}_{0.09}$ & 1.63 $\pm^{0.16}_{0.15}$ & 0.104 $\pm^{0.005}_{0.004}$ &267.3 $\pm^{10.7}_{10.5}$ & 5.49 $\pm^{0.31}_{0.29}$ & 2.78 $\pm^{0.35}_{0.31}$ & 0.198 $\pm^{0.014}_{0.012}$\\
\underline{A1576}\\
& Vikh & 114.8 $\pm^{8.5}_{9.8}$ & 2.23 $\pm^{0.28}_{0.29}$ & 2.15 $\pm^{0.52}_{0.50}$ & 0.104 $\pm^{0.014}_{0.010}$ &239.6 $\pm^{20.2}_{22.0}$ & 7.01 $\pm^{0.78}_{0.79}$ & 3.91 $\pm^{1.07}_{0.98}$ & 0.179 $\pm^{0.033}_{0.024}$\\ 
& Poly & 114.5 $\pm^{6.5}_{6.7}$ & 2.28 $\pm^{0.22}_{0.21}$ & 2.14 $\pm^{0.38}_{0.35}$ & 0.107 $\pm^{0.010}_{0.008}$ &242.5 $\pm^{15.8}_{15.6}$ & 7.26 $\pm^{0.56}_{0.57}$ & 4.05 $\pm^{0.84}_{0.73}$ & 0.179 $\pm^{0.024}_{0.020}$\\
\underline{A1682}\\
& Vikh & 127.4 $\pm^{10.1}_{11.0}$ & 1.66 $\pm^{0.26}_{0.27}$ & 1.75 $\pm^{0.45}_{0.42}$ & 0.095 $\pm^{0.010}_{0.008}$ &274.7 $\pm^{24.0}_{23.0}$ & 6.41 $\pm^{0.74}_{0.71}$ & 3.50 $\pm^{1.00}_{0.81}$ & 0.183 $\pm^{0.029}_{0.024}$\\ 
& Poly & 122.0 $\pm^{6.5}_{6.6}$ & 1.54 $\pm^{0.16}_{0.16}$ & 1.53 $\pm^{0.26}_{0.23}$ & 0.101 $\pm^{0.006}_{0.006}$ &299.4 $\pm^{16.9}_{18.2}$ & 7.18 $\pm^{0.53}_{0.59}$ & 4.53 $\pm^{0.81}_{0.78}$ & 0.159 $\pm^{0.017}_{0.014}$\\
\underline{A1758}\\
& Vikh & 130.7 $\pm^{20.3}_{22.5}$ & 3.12 $\pm^{0.96}_{0.96}$ & 3.17 $\pm^{1.72}_{1.37}$ & 0.098 $\pm^{0.022}_{0.015}$ &334.8 $\pm^{39.9}_{42.6}$ & 13.84 $\pm^{2.25}_{2.22}$ & 10.66 $\pm^{4.28}_{3.57}$ & 0.130 $\pm^{0.034}_{0.023}$\\ 
& Poly & 130.3 $\pm^{10.2}_{18.6}$ & 3.09 $\pm^{0.49}_{0.80}$ & 3.14 $\pm^{0.79}_{1.16}$ & 0.098 $\pm^{0.017}_{0.008}$ &370.1 $\pm^{21.8}_{42.5}$ & 14.84 $\pm^{0.92}_{1.75}$ & 14.41 $\pm^{2.70}_{4.42}$ & 0.104 $\pm^{0.028}_{0.013}$\\
\underline{A1763}\\
& Vikh & 140.0 $\pm^{8.9}_{9.0}$ & 2.65 $\pm^{0.31}_{0.28}$ & 2.24 $\pm^{0.46}_{0.40}$ & 0.119 $\pm^{0.011}_{0.009}$ &348.6 $\pm^{25.3}_{24.7}$ & 11.65 $\pm^{1.12}_{1.10}$ & 6.91 $\pm^{1.61}_{1.37}$ & 0.169 $\pm^{0.022}_{0.019}$\\ 
& Poly & 139.7 $\pm^{6.8}_{4.6}$ & 2.68 $\pm^{0.23}_{0.15}$ & 2.22 $\pm^{0.34}_{0.21}$ & 0.120 $\pm^{0.005}_{0.007}$ &330.0 $\pm^{13.6}_{10.3}$ & 10.72 $\pm^{0.57}_{0.43}$ & 5.86 $\pm^{0.75}_{0.53}$ & 0.183 $\pm^{0.010}_{0.012}$\\
\underline{A1835}\\
& Vikh & 161.7 $\pm^{6.5}_{7.2}$ & 4.60 $\pm^{0.23}_{0.25}$ & 4.74 $\pm^{0.60}_{0.61}$ & 0.097 $\pm^{0.008}_{0.007}$ &323.2 $\pm^{8.6}_{8.6}$ & 10.75 $\pm^{0.28}_{0.29}$ & 7.56 $\pm^{0.62}_{0.59}$ & 0.142 $\pm^{0.008}_{0.007}$\\ 
& Poly & 143.8 $\pm^{2.8}_{2.9}$ & 4.00 $\pm^{0.10}_{0.10}$ & 3.33 $\pm^{0.20}_{0.20}$ & 0.120 $\pm^{0.004}_{0.004}$ &309.6 $\pm^{8.2}_{8.0}$ & 10.36 $\pm^{0.27}_{0.27}$ & 6.65 $\pm^{0.54}_{0.50}$ & 0.156 $\pm^{0.008}_{0.008}$\\
\underline{A1914}\\
& Vikh & 216.7 $\pm^{7.5}_{7.5}$ & 4.71 $\pm^{0.21}_{0.21}$ & 4.22 $\pm^{0.45}_{0.42}$ & 0.112 $\pm^{0.007}_{0.007}$ &444.0 $\pm^{23.2}_{25.6}$ & 10.74 $\pm^{0.56}_{0.61}$ & 7.25 $\pm^{1.20}_{1.19}$ & 0.148 $\pm^{0.019}_{0.015}$\\ 
& Poly & 219.4 $\pm^{7.7}_{7.8}$ & 4.82 $\pm^{0.25}_{0.25}$ & 4.37 $\pm^{0.48}_{0.45}$ & 0.110 $\pm^{0.007}_{0.006}$ &487.5 $\pm^{32.7}_{35.5}$ & 11.30 $\pm^{0.58}_{0.60}$ & 9.59 $\pm^{2.07}_{1.94}$ & 0.118 $\pm^{0.022}_{0.017}$\\
\underline{A2111}\\
& Vikh & 129.9 $\pm^{10.8}_{8.7}$ & 1.96 $\pm^{0.28}_{0.23}$ & 1.91 $\pm^{0.51}_{0.36}$ & 0.103 $\pm^{0.009}_{0.010}$ &296.8 $\pm^{19.5}_{17.0}$ & 7.68 $\pm^{0.69}_{0.62}$ & 4.56 $\pm^{0.96}_{0.74}$ & 0.168 $\pm^{0.017}_{0.017}$\\ 
& Poly & 121.1 $\pm^{4.9}_{5.0}$ & 1.74 $\pm^{0.13}_{0.13}$ & 1.55 $\pm^{0.20}_{0.18}$ & 0.112 $\pm^{0.005}_{0.005}$ &305.3 $\pm^{11.9}_{13.2}$ & 8.04 $\pm^{0.42}_{0.48}$ & 4.96 $\pm^{0.60}_{0.61}$ & 0.162 $\pm^{0.012}_{0.010}$\\ 
\underline{A2204}\\
& Vikh & 244.9 $\pm^{6.1}_{6.0}$ & 4.75 $\pm^{0.15}_{0.15}$ & 4.45 $\pm^{0.34}_{0.32}$ & 0.107 $\pm^{0.005}_{0.004}$ &486.6 $\pm^{16.2}_{15.6}$ & 11.36 $\pm^{0.41}_{0.39}$ & 6.98 $\pm^{0.72}_{0.65}$ & 0.163 $\pm^{0.010}_{0.010}$\\ 
& Poly & 233.7 $\pm^{4.0}_{4.0}$ & 4.50 $\pm^{0.10}_{0.10}$ & 3.87 $\pm^{0.20}_{0.20}$ & 0.116 $\pm^{0.004}_{0.003}$ &496.3 $\pm^{11.9}_{10.4}$ & 11.58 $\pm^{0.30}_{0.27}$ & 7.40 $\pm^{0.55}_{0.46}$ & 0.156 $\pm^{0.007}_{0.007}$\\ 
\underline{A2219}\\
& Vikh & 169.3 $\pm^{6.4}_{6.6}$ & 5.70 $\pm^{0.39}_{0.39}$ & 4.07 $\pm^{0.48}_{0.46}$ & 0.140 $\pm^{0.007}_{0.007}$ &348.6 $\pm^{18.0}_{18.6}$ & 16.77 $\pm^{1.17}_{1.17}$ & 7.11 $\pm^{1.16}_{1.08}$ & 0.236 $\pm^{0.023}_{0.019}$\\ 
& Poly & 166.2 $\pm^{5.0}_{5.3}$ & 5.52 $\pm^{0.32}_{0.32}$ & 3.85 $\pm^{0.36}_{0.35}$ & 0.143 $\pm^{0.006}_{0.005}$ &369.6 $\pm^{16.8}_{16.6}$ & 18.21 $\pm^{1.02}_{1.00}$ & 8.48 $\pm^{1.21}_{1.09}$ & 0.215 $\pm^{0.018}_{0.017}$\\ 
\underline{A2261}\\
& Vikh & 154.5 $\pm^{7.7}_{6.7}$ & 3.63 $\pm^{0.25}_{0.21}$ & 3.04 $\pm^{0.48}_{0.38}$ & 0.119 $\pm^{0.009}_{0.009}$ &319.3 $\pm^{18.8}_{23.4}$ & 9.79 $\pm^{0.65}_{0.79}$ & 5.37 $\pm^{1.01}_{1.10}$ & 0.182 $\pm^{0.028}_{0.019}$\\ 
& Poly & 144.4 $\pm^{5.4}_{5.3}$ & 3.32 $\pm^{0.17}_{0.16}$ & 2.48 $\pm^{0.29}_{0.26}$ & 0.134 $\pm^{0.008}_{0.008}$ &308.0 $\pm^{13.7}_{14.4}$ & 9.37 $\pm^{0.48}_{0.50}$ & 4.82 $\pm^{0.67}_{0.65}$ & 0.194 $\pm^{0.018}_{0.015}$\\ 
\underline{A2390}\\
& Vikh & 188.7 $\pm^{12.1}_{11.8}$ & 5.71 $\pm^{0.57}_{0.54}$ & 6.11 $\pm^{1.25}_{1.07}$ & 0.093 $\pm^{0.009}_{0.008}$ &387.7 $\pm^{34.0}_{35.3}$ & 13.84 $\pm^{1.21}_{1.24}$ & 10.60 $\pm^{3.04}_{2.64}$ & 0.131 $\pm^{0.028}_{0.020}$\\ 
& Poly & 180.2 $\pm^{7.2}_{7.3}$ & 5.33 $\pm^{0.35}_{0.33}$ & 5.32 $\pm^{0.66}_{0.62}$ & 0.100 $\pm^{0.006}_{0.006}$ &372.0 $\pm^{24.6}_{21.6}$ & 13.33 $\pm^{0.84}_{0.76}$ & 9.36 $\pm^{1.98}_{1.54}$ & 0.142 $\pm^{0.018}_{0.018}$\\ 
\underline{A2552}\\
& Vikh & 115.9 $\pm^{8.1}_{9.5}$ & 3.05 $\pm^{0.35}_{0.39}$ & 2.67 $\pm^{0.60}_{0.60}$ & 0.114 $\pm^{0.014}_{0.011}$ &247.2 $\pm^{13.7}_{16.4}$ & 9.66 $\pm^{0.70}_{0.81}$ & 5.19 $\pm^{0.91}_{0.96}$ & 0.186 $\pm^{0.023}_{0.017}$\\ 
& Poly & 109.5 $\pm^{4.6}_{6.0}$ & 2.80 $\pm^{0.19}_{0.25}$ & 2.25 $\pm^{0.29}_{0.35}$ & 0.124 $\pm^{0.010}_{0.007}$ &254.2 $\pm^{10.2}_{13.3}$ & 9.97 $\pm^{0.53}_{0.66}$ & 5.64 $\pm^{0.71}_{0.84}$ & 0.177 $\pm^{0.017}_{0.012}$\\ 
\underline{A2631}\\
& Vikh & 117.7 $\pm^{7.5}_{8.7}$ & 2.87 $\pm^{0.33}_{0.36}$ & 2.30 $\pm^{0.47}_{0.47}$ & 0.125 $\pm^{0.012}_{0.009}$ &273.4 $\pm^{17.9}_{20.3}$ & 10.13 $\pm^{0.67}_{0.78}$ & 5.75 $\pm^{1.21}_{1.18}$ & 0.176 $\pm^{0.029}_{0.021}$\\ 
& Poly & 117.4 $\pm^{6.5}_{6.7}$ & 2.87 $\pm^{0.28}_{0.29}$ & 2.28 $\pm^{0.40}_{0.37}$ & 0.126 $\pm^{0.009}_{0.008}$ &282.0 $\pm^{21.2}_{17.8}$ & 10.50 $\pm^{0.67}_{0.62}$ & 6.31 $\pm^{1.54}_{1.12}$ & 0.166 $\pm^{0.024}_{0.024}$\\ 
\underline{A267}\\
& Vikh & 115.4 $\pm^{9.3}_{10.4}$ & 1.61 $\pm^{0.19}_{0.19}$ & 1.37 $\pm^{0.36}_{0.34}$ & 0.117 $\pm^{0.020}_{0.014}$ &251.5 $\pm^{16.9}_{18.6}$ & 4.82 $\pm^{0.33}_{0.34}$ & 2.83 $\pm^{0.61}_{0.58}$ & 0.170 $\pm^{0.029}_{0.021}$\\ 
& Poly & 108.5 $\pm^{4.3}_{4.7}$ & 1.49 $\pm^{0.09}_{0.10}$ & 1.14 $\pm^{0.14}_{0.14}$ & 0.131 $\pm^{0.009}_{0.008}$ &224.3 $\pm^{13.9}_{13.5}$ & 4.36 $\pm^{0.27}_{0.27}$ & 2.01 $\pm^{0.40}_{0.34}$ & 0.217 $\pm^{0.029}_{0.025}$\\ 
\underline{A520}\\
& Vikh & 166.5 $\pm^{6.0}_{7.2}$ & 3.16 $\pm^{0.22}_{0.25}$ & 2.82 $\pm^{0.31}_{0.35}$ & 0.112 $\pm^{0.006}_{0.005}$ &367.1 $\pm^{9.3}_{9.2}$ & 10.71 $\pm^{0.30}_{0.29}$ & 6.04 $\pm^{0.47}_{0.44}$ & 0.177 $\pm^{0.009}_{0.009}$\\ 
& Poly & 128.9 $\pm^{4.3}_{4.4}$ & 1.83 $\pm^{0.12}_{0.13}$ & 1.31 $\pm^{0.13}_{0.13}$ & 0.140 $\pm^{0.005}_{0.005}$ &393.9 $\pm^{11.7}_{12.0}$ & 11.90 $\pm^{0.36}_{0.39}$ & 7.47 $\pm^{0.69}_{0.66}$ & 0.159 $\pm^{0.010}_{0.009}$\\ 
\underline{A586}\\
& Vikh & 176.8 $\pm^{4.0}_{4.2}$ & 2.26 $\pm^{0.06}_{0.07}$ & 2.28 $\pm^{0.16}_{0.16}$ & 0.099 $\pm^{0.004}_{0.004}$ &354.3 $\pm^{17.3}_{19.4}$ & 5.69 $\pm^{0.29}_{0.33}$ & 3.67 $\pm^{0.56}_{0.57}$ & 0.155 $\pm^{0.018}_{0.014}$\\ 
& Poly & 177.9 $\pm^{5.5}_{5.8}$ & 2.30 $\pm^{0.09}_{0.09}$ & 2.32 $\pm^{0.22}_{0.22}$ & 0.099 $\pm^{0.006}_{0.005}$ &357.1 $\pm^{14.7}_{15.7}$ & 5.66 $\pm^{0.24}_{0.24}$ & 3.76 $\pm^{0.48}_{0.47}$ & 0.151 $\pm^{0.014}_{0.012}$\\ 
\underline{A611}\\
& Vikh & 125.3 $\pm^{8.3}_{10.1}$ & 2.79 $\pm^{0.26}_{0.29}$ & 3.02 $\pm^{0.64}_{0.67}$ & 0.092 $\pm^{0.014}_{0.009}$ &285.3 $\pm^{22.2}_{20.9}$ & 8.36 $\pm^{0.52}_{0.60}$ & 7.13 $\pm^{1.80}_{1.45}$ & 0.118 $\pm^{0.019}_{0.019}$\\ 
& Poly & 124.9 $\pm^{8.2}_{7.4}$ & 2.81 $\pm^{0.25}_{0.22}$ & 2.99 $\pm^{0.63}_{0.50}$ & 0.094 $\pm^{0.010}_{0.010}$ &269.3 $\pm^{27.0}_{28.0}$ & 7.71 $\pm^{0.73}_{0.83}$ & 6.00 $\pm^{1.99}_{1.69}$ & 0.128 $\pm^{0.032}_{0.023}$\\ 
\underline{A665}\\
& Vikh & 166.4 $\pm^{11.5}_{8.0}$ & 2.22 $\pm^{0.29}_{0.19}$ & 2.23 $\pm^{0.50}_{0.31}$ & 0.100 $\pm^{0.006}_{0.008}$ &464.9 $\pm^{29.8}_{33.6}$ & 11.82 $\pm^{0.87}_{1.02}$ & 9.74 $\pm^{1.99}_{1.96}$ & 0.121 $\pm^{0.017}_{0.013}$\\ 
& Poly & 178.5 $\pm^{5.1}_{6.2}$ & 2.56 $\pm^{0.13}_{0.16}$ & 2.76 $\pm^{0.24}_{0.28}$ & 0.093 $\pm^{0.004}_{0.003}$ &412.7 $\pm^{16.9}_{15.6}$ & 9.87 $\pm^{0.55}_{0.51}$ & 6.81 $\pm^{0.87}_{0.74}$ & 0.145 $\pm^{0.009}_{0.009}$\\ 
\underline{A68}\\
& Vikh & 137.7 $\pm^{10.7}_{12.0}$ & 3.09 $\pm^{0.36}_{0.39}$ & 2.96 $\pm^{0.75}_{0.71}$ & 0.105 $\pm^{0.016}_{0.012}$ &305.3 $\pm^{38.0}_{33.8}$ & 8.31 $\pm^{0.83}_{0.81}$ & 6.46 $\pm^{2.73}_{1.91}$ & 0.129 $\pm^{0.037}_{0.031}$\\ 
& Poly & 141.2 $\pm^{8.6}_{8.6}$ & 3.24 $\pm^{0.30}_{0.30}$ & 3.20 $\pm^{0.62}_{0.55}$ & 0.101 $\pm^{0.010}_{0.009}$ &290.5 $\pm^{32.1}_{25.3}$ & 7.92 $\pm^{0.60}_{0.55}$ & 5.57 $\pm^{2.06}_{1.33}$ & 0.143 $\pm^{0.032}_{0.032}$\\ 
\underline{A697}\\
& Vikh & 135.3 $\pm^{10.4}_{11.1}$ & 4.58 $\pm^{0.59}_{0.62}$ & 3.61 $\pm^{0.90}_{0.82}$ & 0.127 $\pm^{0.015}_{0.013}$ &309.9 $\pm^{26.5}_{30.2}$ & 15.48 $\pm^{1.68}_{1.83}$ & 8.68 $\pm^{2.43}_{2.30}$ & 0.178 $\pm^{0.035}_{0.024}$\\ 
& Poly & 130.7 $\pm^{6.7}_{7.8}$ & 4.32 $\pm^{0.38}_{0.44}$ & 3.26 $\pm^{0.53}_{0.55}$ & 0.133 $\pm^{0.011}_{0.009}$ &326.8 $\pm^{20.4}_{18.3}$ & 16.60 $\pm^{1.22}_{1.12}$ & 10.18 $\pm^{2.03}_{1.62}$ & 0.163 $\pm^{0.019}_{0.018}$\\ 
\underline{A773}\\
& Vikh & 144.0 $\pm^{7.5}_{6.7}$ & 2.71 $\pm^{0.20}_{0.19}$ & 2.27 $\pm^{0.38}_{0.30}$ & 0.119 $\pm^{0.009}_{0.009}$ &319.8 $\pm^{14.8}_{13.4}$ & 8.60 $\pm^{0.48}_{0.44}$ & 4.98 $\pm^{0.73}_{0.60}$ & 0.173 $\pm^{0.014}_{0.014}$\\ 
& Poly & 146.1 $\pm^{3.0}_{3.6}$ & 2.80 $\pm^{0.09}_{0.11}$ & 2.38 $\pm^{0.15}_{0.17}$ & 0.118 $\pm^{0.004}_{0.004}$ &334.0 $\pm^{6.6}_{7.7}$ & 9.13 $\pm^{0.19}_{0.23}$ & 5.67 $\pm^{0.35}_{0.38}$ & 0.161 $\pm^{0.007}_{0.006}$\\ 
\underline{A781}\\
& Vikh & 69.7 $\pm^{18.2}_{21.7}$ & 0.71 $\pm^{0.54}_{0.44}$ & 0.57 $\pm^{0.57}_{0.38}$ & 0.126 $\pm^{0.020}_{0.016}$ &249.0 $\pm^{22.7}_{25.0}$ & 8.50 $\pm^{0.90}_{0.98}$ & 5.16 $\pm^{1.55}_{1.40}$ & 0.165 $\pm^{0.036}_{0.025}$\\ 
& Poly & 68.4 $\pm^{8.8}_{8.5}$ & 0.68 $\pm^{0.23}_{0.19}$ & 0.53 $\pm^{0.23}_{0.18}$ & 0.127 $\pm^{0.011}_{0.010}$ &280.4 $\pm^{32.0}_{30.0}$ & 9.95 $\pm^{1.11}_{1.16}$ & 7.37 $\pm^{2.82}_{2.13}$ & 0.135 $\pm^{0.033}_{0.027}$\\ 
\underline{A963}\\
& Vikh & 153.8 $\pm^{8.2}_{9.4}$ & 2.66 $\pm^{0.20}_{0.23}$ & 2.43 $\pm^{0.41}_{0.42}$ & 0.110 $\pm^{0.012}_{0.009}$ &308.2 $\pm^{18.9}_{24.1}$ & 6.49 $\pm^{0.34}_{0.46}$ & 3.91 $\pm^{0.76}_{0.85}$ & 0.166 $\pm^{0.031}_{0.021}$\\ 
& Poly & 149.9 $\pm^{5.0}_{5.3}$ & 2.55 $\pm^{0.12}_{0.12}$ & 2.25 $\pm^{0.23}_{0.23}$ & 0.114 $\pm^{0.007}_{0.006}$ &309.5 $\pm^{17.7}_{16.3}$ & 6.57 $\pm^{0.35}_{0.34}$ & 3.96 $\pm^{0.72}_{0.59}$ & 0.166 $\pm^{0.019}_{0.019}$\\
\\
\end{tabular}
\\\hspace{-50.5em}Continued on next page...
\label{tab:masses}
\end{table*}

\begin{table*}\scriptsize
\renewcommand\thetable{\ref{tab:masses}}
\centering
\caption{Cluster Masses--Continued}
\begin{tabular}{lcccccccccc}
\hline
\hline
& & \multicolumn{4}{c}{$\Delta = 2500$} & \multicolumn{4}{c}{$\Delta = 500$}\\
Cluster & Model & $r_{\Delta}$ & $M_{\textrm{gas}}$ & $M_{\textrm{tot}}$ & $f_{\textrm{gas}}$ & $r_{\Delta}$ & $M_{\textrm{gas}}$ & $M_{\textrm{tot}}$ & $f_{\textrm{gas}}$\\
& & (arcsec) & (10$^{13} \,\textrm{M}_{\odot}$) & (10$^{14} \,\textrm{M}_{\odot}$) & & (arcsec) & (10$^{13} \,\textrm{M}_{\odot}$) & (10$^{14} \,\textrm{M}_{\odot}$) &\\
\hline
\underline{M1455}\\
& Vikh & 104.9 $\pm^{3.7}_{4.2}$ & 1.69 $\pm^{0.07}_{0.08}$ & 1.35 $\pm^{0.15}_{0.15}$ & 0.125 $\pm^{0.010}_{0.008}$ &236.8 $\pm^{9.0}_{10.1}$ & 4.64 $\pm^{0.18}_{0.21}$ & 3.11 $\pm^{0.37}_{0.38}$ & 0.149 $\pm^{0.013}_{0.011}$\\ 
& Poly & 105.3 $\pm^{2.5}_{2.0}$ & 1.70 $\pm^{0.04}_{0.03}$ & 1.37 $\pm^{0.10}_{0.08}$ & 0.124 $\pm^{0.005}_{0.005}$ &212.9 $\pm^{6.6}_{5.0}$ & 4.24 $\pm^{0.12}_{0.09}$ & 2.26 $\pm^{0.22}_{0.16}$ & 0.187 $\pm^{0.009}_{0.011}$\\ 
\underline{R0437}\\
& Vikh & 124.1 $\pm^{7.9}_{8.1}$ & 2.80 $\pm^{0.25}_{0.25}$ & 2.86 $\pm^{0.58}_{0.52}$ & 0.098 $\pm^{0.011}_{0.009}$ &262.9 $\pm^{24.7}_{25.6}$ & 7.59 $\pm^{0.81}_{0.76}$ & 5.44 $\pm^{1.68}_{1.44}$ & 0.140 $\pm^{0.031}_{0.022}$\\ 
& Poly & 116.8 $\pm^{3.8}_{3.5}$ & 2.59 $\pm^{0.12}_{0.11}$ & 2.38 $\pm^{0.24}_{0.21}$ & 0.108 $\pm^{0.006}_{0.005}$ &279.8 $\pm^{10.6}_{10.1}$ & 8.01 $\pm^{0.32}_{0.30}$ & 6.55 $\pm^{0.77}_{0.68}$ & 0.122 $\pm^{0.010}_{0.009}$\\ 
\underline{R0439}\\
& Vikh & 146.4 $\pm^{8.5}_{8.6}$ & 2.94 $\pm^{0.24}_{0.25}$ & 2.77 $\pm^{0.51}_{0.46}$ & 0.106 $\pm^{0.011}_{0.009}$ &285.1 $\pm^{23.0}_{24.4}$ & 6.95 $\pm^{0.50}_{0.55}$ & 4.09 $\pm^{1.07}_{0.96}$ & 0.170 $\pm^{0.035}_{0.026}$\\ 
& Poly & 138.3 $\pm^{5.8}_{5.8}$ & 2.72 $\pm^{0.17}_{0.17}$ & 2.33 $\pm^{0.30}_{0.28}$ & 0.117 $\pm^{0.008}_{0.008}$ &278.6 $\pm^{18.8}_{17.0}$ & 6.80 $\pm^{0.40}_{0.40}$ & 3.81 $\pm^{0.82}_{0.65}$ & 0.178 $\pm^{0.025}_{0.024}$\\ 
\underline{R1720}\\
& Vikh & 213.3 $\pm^{11.2}_{7.2}$ & 3.31 $\pm^{0.23}_{0.15}$ & 3.59 $\pm^{0.60}_{0.35}$ & 0.092 $\pm^{0.005}_{0.008}$ &442.5 $\pm^{41.0}_{22.2}$ & 8.23 $\pm^{0.86}_{0.44}$ & 6.41 $\pm^{1.95}_{0.92}$ & 0.128 $\pm^{0.014}_{0.020}$\\ 
& Poly & 189.1 $\pm^{5.4}_{5.1}$ & 2.84 $\pm^{0.11}_{0.09}$ & 2.50 $\pm^{0.22}_{0.20}$ & 0.113 $\pm^{0.006}_{0.005}$ &427.7 $\pm^{12.5}_{12.8}$ & 7.91 $\pm^{0.22}_{0.21}$ & 5.79 $\pm^{0.52}_{0.51}$ & 0.137 $\pm^{0.009}_{0.008}$\\ 
\underline{R2129}\\
& Vikh & 143.9 $\pm^{8.1}_{7.5}$ & 3.02 $\pm^{0.23}_{0.21}$ & 2.78 $\pm^{0.49}_{0.41}$ & 0.109 $\pm^{0.010}_{0.010}$ &283.9 $\pm^{21.0}_{18.1}$ & 7.41 $\pm^{0.58}_{0.51}$ & 4.26 $\pm^{1.02}_{0.76}$ & 0.174 $\pm^{0.024}_{0.023}$\\ 
& Poly & 138.4 $\pm^{4.8}_{4.9}$ & 2.87 $\pm^{0.13}_{0.13}$ & 2.46 $\pm^{0.27}_{0.25}$ & 0.116 $\pm^{0.007}_{0.007}$ &292.4 $\pm^{13.6}_{14.1}$ & 7.66 $\pm^{0.38}_{0.38}$ & 4.65 $\pm^{0.68}_{0.64}$ & 0.165 $\pm^{0.017}_{0.014}$\\ 
\underline{Z2089}\\
& Vikh & 91.1 $\pm^{5.6}_{5.4}$ & 1.54 $\pm^{0.12}_{0.11}$ & 1.07 $\pm^{0.21}_{0.18}$ & 0.143 $\pm^{0.017}_{0.014}$ &211.0 $\pm^{14.7}_{16.9}$ & 4.38 $\pm^{0.31}_{0.33}$ & 2.67 $\pm^{0.60}_{0.59}$ & 0.165 $\pm^{0.032}_{0.023}$\\ 
& Poly & 107.7 $\pm^{4.2}_{4.2}$ & 1.40 $\pm^{0.07}_{0.06}$ & 1.16 $\pm^{0.14}_{0.13}$ & 0.121 $\pm^{0.010}_{0.009}$ &225.9 $\pm^{12.1}_{14.1}$ & 3.48 $\pm^{0.19}_{0.20}$ & 2.14 $\pm^{0.36}_{0.38}$ & 0.163 $\pm^{0.024}_{0.018}$\\ 
\underline{Z3146}\\
& Vikh & 132.8 $\pm^{8.5}_{8.0}$ & 4.82 $\pm^{0.37}_{0.36}$ & 3.68 $\pm^{0.75}_{0.63}$ & 0.131 $\pm^{0.015}_{0.014}$ &277.2 $\pm^{20.9}_{23.1}$ & 10.96 $\pm^{0.70}_{0.78}$ & 6.69 $\pm^{1.62}_{1.54}$ & 0.164 $\pm^{0.034}_{0.024}$\\ 
& Poly & 120.5 $\pm^{3.4}_{3.6}$ & 4.29 $\pm^{0.15}_{0.15}$ & 2.75 $\pm^{0.24}_{0.24}$ & 0.156 $\pm^{0.009}_{0.008}$ &273.6 $\pm^{9.8}_{10.8}$ & 10.79 $\pm^{0.32}_{0.34}$ & 6.42 $\pm^{0.72}_{0.73}$ & 0.168 $\pm^{0.016}_{0.013}$\\ 
\underline{Z5247}\\
& Vikh & 91.9 $\pm^{10.3}_{13.5}$ & 0.61 $\pm^{0.18}_{0.20}$ & 0.69 $\pm^{0.25}_{0.26}$ & 0.089 $\pm^{0.009}_{0.006}$ &222.0 $\pm^{11.5}_{13.5}$ & 4.14 $\pm^{0.36}_{0.40}$ & 1.93 $\pm^{0.32}_{0.33}$ & 0.215 $\pm^{0.021}_{0.016}$\\ 
& Poly & 74.6 $\pm^{6.0}_{5.4}$ & 0.36 $\pm^{0.08}_{0.06}$ & 0.37 $\pm^{0.09}_{0.07}$ & 0.099 $\pm^{0.004}_{0.004}$ &255.2 $\pm^{14.0}_{13.3}$ & 5.56 $\pm^{0.46}_{0.45}$ & 2.93 $\pm^{0.51}_{0.43}$ & 0.190 $\pm^{0.016}_{0.016}$\\ 
\underline{Z5768}\\
& Vikh & 81.4 $\pm^{10.3}_{11.7}$ & 0.47 $\pm^{0.10}_{0.10}$ & 0.68 $\pm^{0.29}_{0.25}$ & 0.069 $\pm^{0.018}_{0.012}$ &178.1 $\pm^{21.0}_{24.0}$ & 1.85 $\pm^{0.30}_{0.33}$ & 1.43 $\pm^{0.57}_{0.50}$ & 0.131 $\pm^{0.036}_{0.025}$\\ 
& Poly & 80.6 $\pm^{10.6}_{9.4}$ & 0.54 $\pm^{0.12}_{0.11}$ & 0.66 $\pm^{0.30}_{0.21}$ & 0.080 $\pm^{0.014}_{0.012}$ &191.8 $\pm^{18.9}_{17.9}$ & 2.26 $\pm^{0.27}_{0.28}$ & 1.78 $\pm^{0.58}_{0.45}$ & 0.128 $\pm^{0.029}_{0.025}$\\ 
\underline{Z7215}\\
& Vikh & 128.6 $\pm^{16.2}_{15.9}$ & 2.72 $\pm^{0.51}_{0.48}$ & 3.31 $\pm^{1.41}_{1.08}$ & 0.082 $\pm^{0.019}_{0.014}$ &273.7 $\pm^{25.6}_{25.0}$ & 7.10 $\pm^{0.66}_{0.59}$ & 6.39 $\pm^{1.96}_{1.60}$ & 0.112 $\pm^{0.027}_{0.021}$\\ 
& Poly & 114.5 $\pm^{13.1}_{12.4}$ & 2.30 $\pm^{0.41}_{0.37}$ & 2.34 $\pm^{0.89}_{0.68}$ & 0.098 $\pm^{0.019}_{0.015}$ &319.0 $\pm^{44.8}_{45.0}$ & 7.80 $\pm^{0.88}_{0.76}$ & 10.11 $\pm^{4.88}_{3.70}$ & 0.078 $\pm^{0.034}_{0.023}$\\ 

\hline
\end{tabular}
\end{table*}


\clearpage
\newpage

\begin{appendices}
\section{Appendix: Model Fits to Temperature Profiles}
\label{sec:appendixA}
Temperature profiles for all the clusters with the \cite{vikhlinin2006} model (blue)
and the \cite{bulbul2010} model (red) are shown in Figure~\ref{fig:T-fits}.
Table \ref{tab:Textrap} shows the radius out to which the 
temperature was measured and the estimated $r_{500}$ value for
each cluster. Out of the 7 clusters, 5 clusters have temperature profiles which extended to $\geq 85\%$ of $r_{500}$.
The temperature profiles of \textit{Abell~1758} and \textit{Abell~611} only 
reached 73\% and 63\% of $r_{500}$, respectively.
Both of these clusters suffered from solar flares, which did not allow the determination of the temperature profile out to $r_{500}$.
The effects of extrapolating the temperature profiles for these 7 clusters are not believed to significantly affect
mass measurements.

\begin{table*}
\centering
\caption{Extrapolation of Temperature Profiles out to $r_{500}$}
\begin{tabular}{lccc}
\hline
\hline
\multirow{2}{*}{Cluster} & Max Radius of $T(r)$ & $r_{500}$ & Per cent of $r_{500}$\\
                         & (arcsec) & (arcsec) & (\%)\\
\hline
A1758 & $240$ & $330$ & $73$\\
A1914 & $420$ & $440$ & $95$\\
A611  & $180$ & $285$ & $63$\\
RX J0437.1+0043 & $240$ & $260$ & $92$\\
RX J1720.1+2638 & $400$ & $440$ & $91$\\
Z2089 & $180$ & $210$ & $86$\\
Z7215 & $240$ & $275$ & $87$\\
\hline
\end{tabular}
\label{tab:Textrap}
\end{table*}

\begin{figure}
\centering
\includegraphics[angle=-90,width=2in]{./FIG/a115_poly_vikh_temp.ps}
\includegraphics[angle=-90,width=2in]{./FIG/a1423_poly_vikh_temp.ps}
\includegraphics[angle=-90,width=2in]{./FIG/a1576_poly_vikh_temp.ps}
\includegraphics[angle=-90,width=2in]{./FIG/a1682_poly_vikh_temp.ps}
\includegraphics[angle=-90,width=2in]{./FIG/a1758_poly_vikh_temp.ps}
\includegraphics[angle=-90,width=2in]{./FIG/a1763_poly_vikh_temp.ps}
\end{figure}
\begin{figure}
\centering
\includegraphics[angle=-90,width=2in]{./FIG/a1835_poly_vikh_temp.ps}
\includegraphics[angle=-90,width=2in]{./FIG/a1914_poly_vikh_temp.ps}
\includegraphics[angle=-90,width=2in]{./FIG/a2111_poly_vikh_temp.ps}
\includegraphics[angle=-90,width=2in]{./FIG/a2204_poly_vikh_temp.ps}
\includegraphics[angle=-90,width=2in]{./FIG/a2219_poly_vikh_temp.ps}
\includegraphics[angle=-90,width=2in]{./FIG/a2261_poly_vikh_temp.ps}
\end{figure}
\begin{figure}
\centering
\includegraphics[angle=-90,width=2in]{./FIG/a2390_poly_vikh_temp.ps}
\includegraphics[angle=-90,width=2in]{./FIG/a2552_poly_vikh_temp.ps}
\includegraphics[angle=-90,width=2in]{./FIG/a2631_poly_vikh_temp.ps}
\includegraphics[angle=-90,width=2in]{./FIG/a267_poly_vikh_temp.ps}
\includegraphics[angle=-90,width=2in]{./FIG/a520_poly_vikh_temp.ps}
\includegraphics[angle=-90,width=2in]{./FIG/a586_poly_vikh_temp.ps}
\end{figure}
\begin{figure}
\centering
\includegraphics[angle=-90,width=2in]{./FIG/a611_poly_vikh_temp.ps}
\includegraphics[angle=-90,width=2in]{./FIG/a665_poly_vikh_temp.ps}
\includegraphics[angle=-90,width=2in]{./FIG/a68_poly_vikh_temp.ps}
\includegraphics[angle=-90,width=2in]{./FIG/a697_poly_vikh_temp.ps}
\includegraphics[angle=-90,width=2in]{./FIG/a773_poly_vikh_temp.ps}
\includegraphics[angle=-90,width=2in]{./FIG/a781_poly_vikh_temp.ps}
\end{figure}
\begin{figure}
\centering
\includegraphics[angle=-90,width=2in]{./FIG/a963_poly_vikh_temp.ps}
\includegraphics[angle=-90,width=2in]{./FIG/ms1455_poly_vikh_temp.ps}
\includegraphics[angle=-90,width=2in]{./FIG/rxj0437_poly_vikh_temp.ps}
\includegraphics[angle=-90,width=2in]{./FIG/rxj0439_poly_vikh_temp.ps}
\includegraphics[angle=-90,width=2in]{./FIG/rxj1720_poly_vikh_temp.ps}
\includegraphics[angle=-90,width=2in]{./FIG/rxj2129_poly_vikh_temp.ps}
\end{figure}
\begin{figure}
\centering
\includegraphics[angle=-90,width=2in]{./FIG/zw2089_poly_vikh_temp.ps}
\includegraphics[angle=-90,width=2in]{./FIG/zw3146_poly_vikh_temp.ps}
\includegraphics[angle=-90,width=2in]{./FIG/zw5247_poly_vikh_temp.ps}
\includegraphics[angle=-90,width=2in]{./FIG/zw5768_poly_vikh_temp.ps}
\includegraphics[angle=-90,width=2in]{./FIG/zw7215_poly_vikh_temp.ps}
\caption{Temperature profiles for all clusters using the \protect\cite{vikhlinin2006} model (blue)
and the \protect\cite{bulbul2010} model (red). The solid lines show the best-fit values, and
the hatched region is the 68.3\% confidence interval.}
\label{fig:T-fits}
\end{figure}
\end{appendices}

\bibliography{ms}
\bibliographystyle{mn2e}

\label{lastpage}
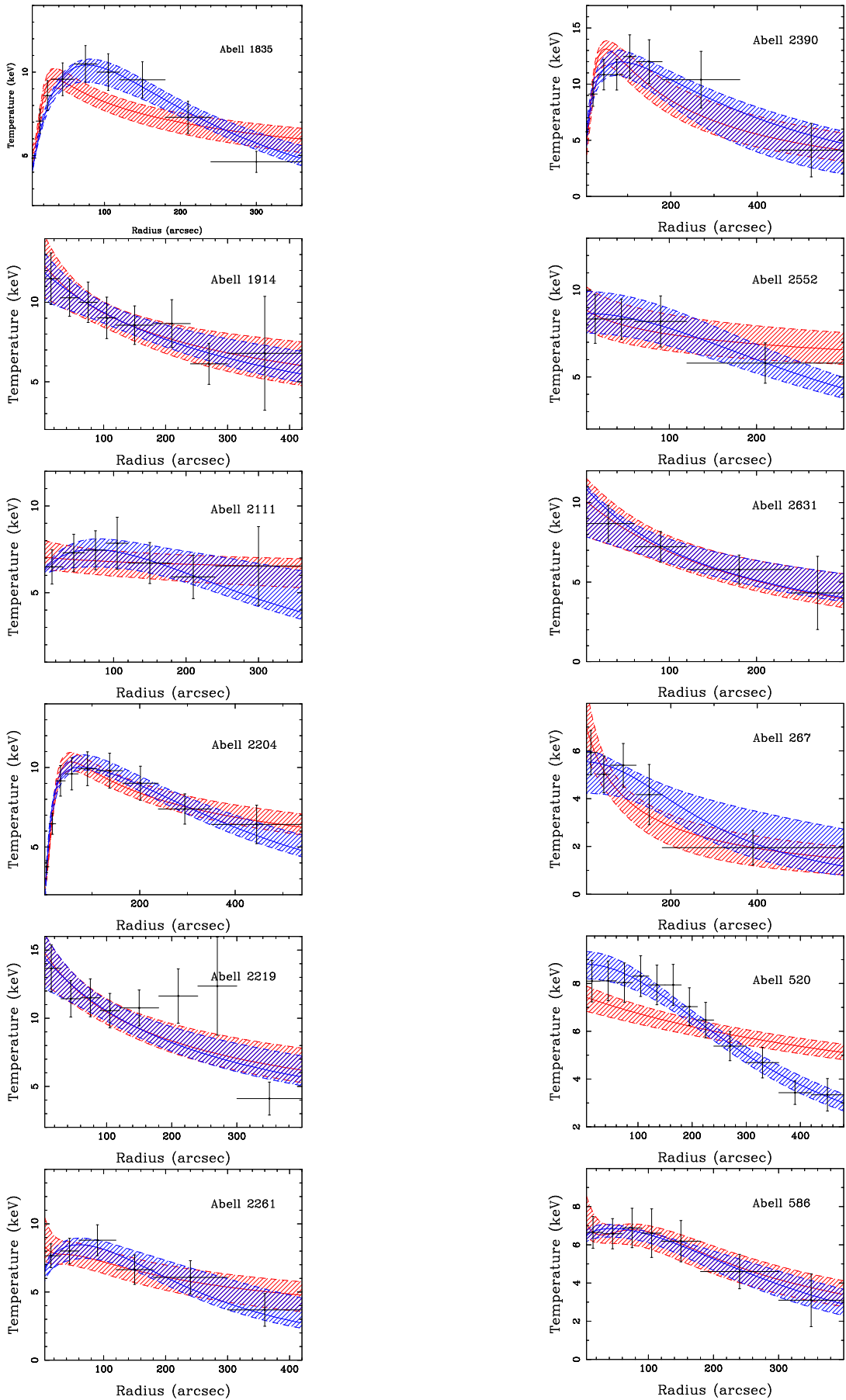
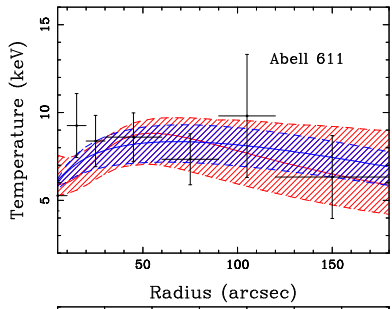
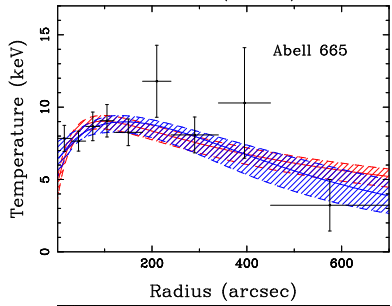
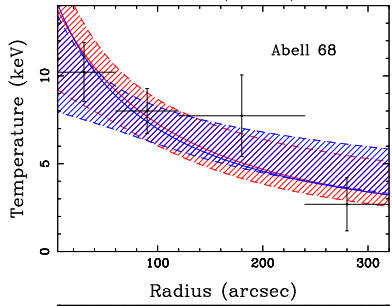
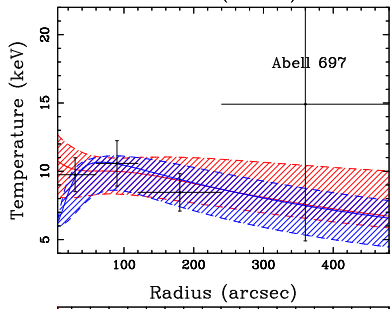
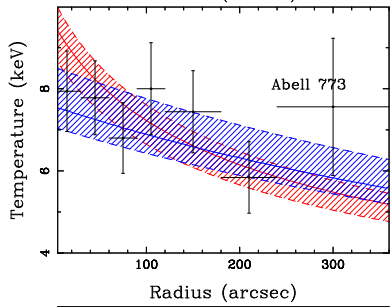
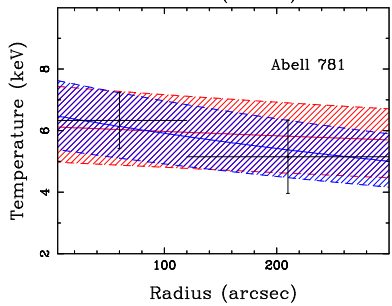
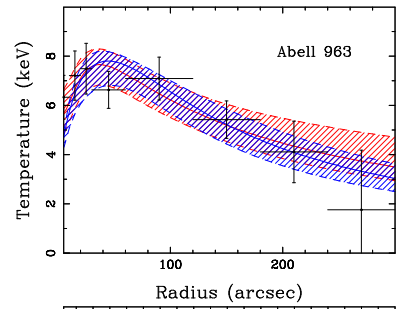
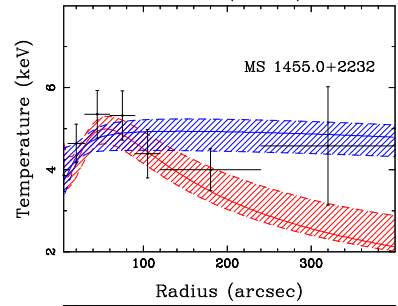
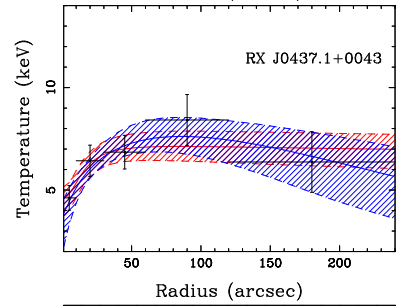
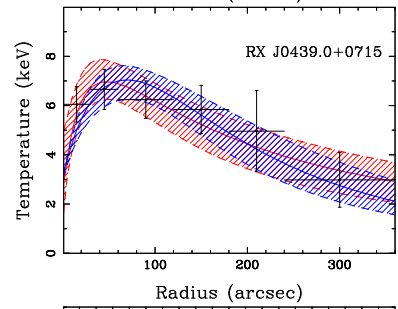
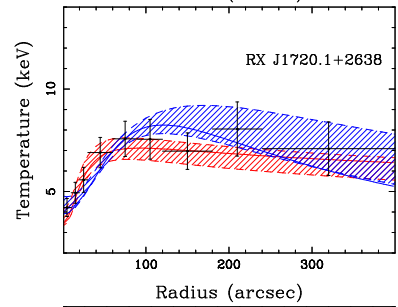
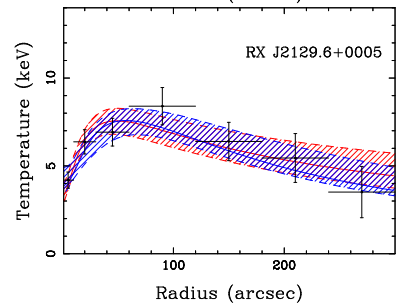
\end{document}